\documentclass[%
 reprint,
 amsmath,amssymb,
 aps,
floatfix,
]{revtex4-2}

\usepackage{graphicx}
\usepackage{dcolumn}
\usepackage{bm}
\usepackage{braket}
\usepackage{mathrsfs}
\usepackage{float}
\usepackage{qcircuit}
\usepackage{tikzit}

\tikzstyle{Z}=[fill=white, draw=black, shape=circle]
\tikzstyle{X}=[fill=black, draw=black, shape=circle]
\tikzstyle{H}=[fill=white, draw=black, shape=rectangle]

\usepackage{tikzscale}
\usepackage[hidelinks]{hyperref}

\def\[#1\]{\begin{align}#1\end{align}}

\def\valem[#1]{\boxed{\textcolor{red}{#1}}}
\def\blue[#1]{\textcolor{blue}{#1}}
\def\green[#1]{\textcolor{green}{#1}}
\def\red[#1]{\textcolor{red}{#1}}
\def\ex[#1]{\exp\left(#1\right)}
\def\ps{\ket{\psi}}

\begin{document}

\preprint{APS/123-QED}

\title{Resource efficient method for representation and measurement of constrained electronic structure states with a quantum computer}

\author{Kaur Kristjuhan}
\author{Mark Nicholas Jones}%
 \email{mark@mqs.dk}
\affiliation{%
Molecular Quantum Solutions ApS, Maskinvej 5, 2860 Søborg, Denmark
}%

\date{\today}

\begin{abstract}
We present a novel method for improving the quantum simulation of the ground state energy of molecules. We perform a pre-processing step classically, which reduces the dimensionality of the problem by generating a custom mapping which excludes states which violate problem constraints. Subsequently, a specialized measurement scheme is used to extract the expectation value of the problem Hamiltonian through this mapping. We demonstrate that this method reduces the amount of quantum resources needed to run a Variational Quantum Eigensolver (VQE) algorithm without making any approximations to the physics of the quantum chemistry problem.
\end{abstract}

\maketitle


\section{Introduction}
The variational method for determining the ground state of a system described by a Hamiltonian $\hat{H}$ entails performing a search for a state $\ps$ which minimizes the expectation value of the Hamiltonian. This relies on the definition of the ground state as the lowest energy eigenstate of the Hamiltonian, meaning that the ground state energy $E_g$ is lower than the expectation value of the Hamiltonian in any other state:
\[
E_g \leq \braket{\psi|\hat{H}|\psi}
\]
In molecular electronic structure problems, the search for the state $\ps$ is performed in a fermionic Fock space $\mathcal{F}$, which is constructed from a finite number of electron orbitals. This search can be performed with the variational quantum eigensolver (VQE). During the execution of the VQE algorithm, a representation of a state $\ps$ is prepared on a quantum computer (QC) and the expectation value of the Hamiltonian is inferred through partial state tomography. A classical optimizer determines which state the QC should prepare in each step of the search, such that after sufficiently many iterations the QC is able to prepare and measure the ground state \cite{peruzzo2014variational,cerezo2021variational,fedorov2022vqe,cao2019quantum,mcardle2020quantum}. There are multiple aspects in which the performance of a VQE can be improved and in this article we will focus on three of those aspects: constraint enforcement, qubit reduction, and efficient measurements.

\subsection{Constraint enforcement}
Often it is desirable to solve an electronic structure problem subject to constraints on certain features such as particle number, spin multiplicity and spatial symmetries. These constraints restrict the set of valid answers to just a subspace of the entire Fock space. It is valuable to develop methods that ensure that the VQE algorithm performs its search within this subspace for various reasons:
\\
\\
1) If a VQE algorithm is allowed to search outside of the desired subspace, it may happen that the algorithm converges to a state outside of the subspace, thereby suggesting an answer which explicitly violates the constraints of the problem and is therefore wrong. Such an answer has very little (if any) utility in solving the original problem and would therefore need to be discarded. Furthermore, for each such occurrence, additional computational processing would need to be done to ensure that the VQE algorithm would not converge to that answer again after a reset.
\\
\\
2) If the problem is constrained to a subspace with lower dimensionality, the search is likely to converge faster and require the optimization of fewer independent parameters.
\\
\\
3) The lowest energy state of a constrained problem might not be the same as the lowest energy state of the unconstrained problem. This means that without explicitly enforcing the constraints, the VQE algorithm may be unable to converge to the desired answer.
\\
\\
Constraints can be enforced by modifying the algorithm designed for solving the unconstrained problem. Known techniques involve modifying either the mapping \cite{gunlycke,fischer2019symmetry, steudtner2018fermion, steudtner2019methods, chamaki2022compact, shee2022qubit}, the Ansatz \cite{romero2018strategies,grimsley2019adaptive,anselmetti2021local, gard2020efficient} or the expectation value \cite{mcclean2016theory,ryabinkin2018constrained, kuroiwa2021penalty}.

\subsubsection{Mapping}
To perform any meaningful quantum computation, problems need to be encoded onto the quantum computer in some manner. In electronic structure problems, the quantum states in the fermionic Fock space $\mathcal{F}$ describe which orbitals are occupied by electrons. For example, the state $\ket{1001}\in \mathcal{F}$ represents a configuration, where the first and the last orbitals are occupied by electrons, whereas the other two are not. The most common and straightforward way of mapping such states onto a quantum computer is by using the Jordan-Wigner transformation \cite{nielsen2005fermionic,jordan1993paulische}, which preserves the notation of these states, while changing their meaning. For example, the state $\ket{1001}\in F$ described earlier is mapped to $\ket{1001}\in\mathcal{H}$, where $\mathcal{H}$ is the Hilbert space of the quantum computer, which has dimension $N=2^Q$, where $Q$ is the number of qubits in the register. In this space, the state $\ket{1001}$ represents the first and last qubits being in the $\ket{1}$ computational basis state, while the others are in the $\ket{0}$ computational basis state. There is no fundamental reason why this particular mapping has to be chosen and many others have been developed \cite{seeley2012bravyi,setia2018bravyi,steudtner2018fermion}. The key insight needed to develop a useful mapping for constraint enforcement is that not all states in $\mathcal{F}$ necessarily need to be mapped to $\mathcal{H}$. That is, if we were able to design a mapping which only maps precisely all those states that satisfy the problem constraints and no others, then we can guarantee that the VQE algorithm is performing its search exclusively among those states. The aspect that makes this approach challenging is that in addition to mapping states, certain Fock space operators must also be mapped. For example, in the Jordan-Wigner encoding, fermionic creation and annihilation operators on $\mathcal{F}$ are mapped to Pauli operators on $\mathcal{H}$ as
\[
\hat{a}_i^{\dagger} &\rightarrow \frac{1}{2} (\hat{X}_i - i\hat{Y}_i) \bigotimes_{j < i} \hat{Z}_j\label{jw1}\\
\hat{a}_i &\rightarrow \frac{1}{2} (\hat{X}_i + i\hat{Y}_i) \bigotimes_{j < i} \hat{Z}_j\label{jw2}
\]
These transformations must be consistent with how the states transform and must preserve the anticommutation relation of fermions:
\[
\{\hat{a}_i,\hat{a}_j\}&=0\\
\{\hat{a}_i^\dagger,\hat{a}_j^\dagger\}&=0\\
\{\hat{a}_i,\hat{a}_j^\dagger\}&=\delta_{ij}
\]
It is not guaranteed that mapping these operators is at all possible when $\mathcal{F}$ is only partially mapped onto $\mathcal{H}$. For example, if we only map states that have a total of two electrons, then neither creation nor annihilation operators can be mapped, because both of them change the particle number of the state. In other words, if either of these operators are applied to a two particle state, the resulting state would no longer have two particles and therefore would not be a state represented in $\mathcal{H}$. Fortunately, it may not be necessary to map these operators at all, given that for the purposes of the VQE algorithm, we only evaluate the expectation value of the Hamiltonian operator $\hat{H}$. As long as $\hat{H}$ can be mapped, the inability to map $\hat{a}$ or $\hat{a}^\dagger$ does not hinder the VQE algorithm. In this example of a two electron constraint, it can be inferred that $\hat{H}$ could in principle be mapped, because the second quantized Hamiltonian of electronic structure problems is built by adding together various terms that all conserve particle number.
Ideas for such mappings has been pursued in references \cite{gunlycke,fischer2019symmetry, steudtner2018fermion, steudtner2019methods, chamaki2022compact, shee2022qubit}.

\subsubsection{Ansatz}
To prepare a state, the QC starts from some initial state $\ket{\psi_0}$, which is operationally simple to prepare. During the quantum computation, a quantum circuit is applied to the initial state to prepare the final state $\ps$. Usually, the initial state is chosen to be the same for all calculations and only the subsequent circuit is modified to produce different states. The effect of the circuit can be expressed as a unitary operator $\hat{U}$, which is dependent on some set of parameters $\boldsymbol{\theta}$. This can be realized in quantum circuits by using parameterized gates, such as rotation gates, where a parameter between 0 and $2\pi$ determines the degree of phase rotation applied by the gate. The classical part (the optimizer) of the VQE algorithm determines the desired values of $\boldsymbol{\theta}$. The resulting state on the QC for given values of $\boldsymbol{\theta}$ is
\[
\ps=\hat{U}(\boldsymbol{\theta})\ket{\psi_0} 
\]
The architecture of the circuit or the functional dependence of $\hat{U}$ on $\boldsymbol{\theta}$ is called the Ansatz. It is not obvious what the best Ansatz is for solving electronic structure problems and many varieties have been developed, guided by a plethora of metrics such as ease of implementation on hardware, number of average iterations required for convergence, depth of quantum circuit needed for implementation, number of multi-qubit gates in the circuit, number of independent parameters $\boldsymbol{\theta}$ and overall simplicity \cite{mezz,romero2018strategies,grimsley2019adaptive,anselmetti2021local, kandala2017hardware}.

Problem constraints can be enforced by starting from an initial state $\ket{\psi_0}$ that respects the constraints and choosing an Ansatz which does not subsequently violate them \cite{gard2020efficient}. Even when a suitable Ansatz is implemented, it is susceptible to two types of errors.

First, precise hardware operation may be impossible due to low gate fidelity. This means that in a real device, the effect of a quantum gate may deviate from its desired effect. As circuits become longer, these effects are compounded, making the implemented Ansatz increasingly different from the designed one, leading to possible constraint violations.

Second, qubit readout errors can lead to inaccurate state tomography. This means that even if the circuit is perfectly executed and the state prepared on the QC adheres to the constraints, faulty readout can cause a constraint violation. For example, in the Jordan-Wigner encoding of molecular Hamiltonians, the logical basis states of qubits represent whether a particular spin orbital is occupied by an electron or not. An inaccurate reading of any one bit would either increase or decrease the number of total electrons by one, which is an obvious constraint violation when trying to simulate a system with a fixed number of electrons. This error may occur regardless of how the Ansatz circuit is chosen.

\subsubsection{Expectation value}
Another approach, independent of the other two, is to add penalty terms to the Hamiltonian \cite{mcclean2016theory,ryabinkin2018constrained, kuroiwa2021penalty}. Instead of finding the state $\ps$ that minimizes $\braket{\psi|\hat{H}|\psi}$, the expression $\braket{\psi|\hat{H}|\psi} + \kappa(\psi)$ could be minimized instead, where $\kappa(\psi)$ represents a function that equals zero when $\ps$ satisfies the constraint and has a positive non-zero value otherwise. For example, the number of electrons in a molecule can be set to two with the following choice:
\[
\kappa(\psi)=w\left(\sum_{i=1}^N\braket{\psi|\hat{a}^\dagger_i \hat{a}_i|\psi}-2\right)^2
\]
where $i$ indexes the $N$ different spin orbitals in the system, $\hat{a}^\dagger$ and $\hat{a}$ are the fermionic creation and annihilation operators and $w$ is an appropriately chosen positive constant that weights the constraint relative to the Hamiltonian and any additional constraints added in this manner. In this approach it is relatively straightforward to add new constraints, each of which may require additional measurements on the quantum computer to evaluate. The choice of $w$ is a separate task that needs to be optimized. A value too small would lack impact on the VQE optimization procedure. A value too large would obstruct the algorithm from converging, because most of the optimization effort would be spent on satisfying the constraints, rather than minimizing the expectation value of the Hamiltonian. The issues are exacerbated when using NISQ hardware, where the estimation of the expectation value is not necessarily accurate. This lack of accuracy can be caused by both systematic errors in state preparation or readout and by statistical errors if not enough measurement repetitions are performed. In the presence of limited accuracy, the penalty terms can never be optimized precisely to zero, creating a perpetual optimization cycle in the case of large $w$.


\subsection{Qubit reduction}
The capabilities of near term Noisy Intermediate Scale Quantum (NISQ) devices are often limited by the number of available qubits. Methods which reduce the number of qubits required to solve a given problem are therefore valuable in multiple respects:
\\
\\
1) Reducing the number of required qubits makes it possible to solve problems on NISQ devices that would otherwise be impossible due to lack of qubits.
\\
\\
2) Algorithms that use more qubits tend to produce more errors when run on NISQ devices due to the involvement of more independently unknown or uncertain hardware parameters. Therefore, algorithms that use fewer qubits are more resilient to errors.
\\
\\
3) Quantum circuits designed for more qubits tend to be deeper, due to the need for including more entangling gates between the qubits. This is especially true for devices which lack full connectivity between qubits, where arbitrary multi-qubit interactions need to be mediated with additional swap gates \cite{o2019generalized}.
\\
\\
Broadly, there are two kinds of techniques to reduce the qubit requirements for electronic structure problems - those which simplify the original problem and those which do not. Often, the various methods are not mutually exclusive and many can be used simultaneously.

\subsubsection{Qubit reduction by problem simplification}
The qubit requirements of quantum simulation of electronic structure problems scale with the number of orbitals that are taken into account in the simulation. Some orbitals contribute more to the simulation than others, so it is a common strategy to exclude less important orbitals from the simulation and replace them with approximations. For example, almost all atoms have their two innermost spin orbitals always occupied in the ground state, leading to the widely used frozen core approximation, which excludes those orbitals from the simulation by assuming that they are fully occupied by electrons \cite{koridon2021orbital, tsuchimochi2022adaptive, li2022toward}.

Another, more indirect way of simplification is to opt for the use of smaller basis sets. Currently in the field of quantum simulation of chemistry, it is common for people to use a minimal basis set such as STO-3G, despite their known lack of accuracy \cite{elfving2020will}. There are many other basis sets, known to represent the electronic structure of atoms more accurately, but often these more accurate basis sets include substantially more orbitals, making their use infeasible on current quantum hardware.

Many approaches also exist which aim to produce an effective Hamiltonian which describes a fictitious system that is designed to be simpler while maintaining an energy close to the original \cite{fujii2022deep, kumar2022quantum, bauman2022coupled, dhawan2020dynamical}

\subsubsection{Non-simplifying qubit reduction}
Occasionally, it is possible to perform some mathematical transformations which cast the problem into an equivalent, but simpler form. One example is the partial encoding of the Fock space introduced earlier in the previous section. Another example is the qubit tapering method \cite{taper}, which exploits the symmetries of the Hamiltonian to transform it into a form which does not necessitate the involvement of some qubits in the calculations. It does not modify the original problem, but simply removes the redundancy hidden within the description. A different example is the Quantum Subspace Expansion (QSE) method \cite{mcclean2017hybrid,urbanek2020chemistry}, which does not reduce qubit requirements explicitly, but uses classical post-processing on VQE results to improve their accuracy to an extent that would otherwise be only achievable with more qubits.

\subsection{Efficient measurements}
Partial state tomography involves performing a set of measurements on the QC and then extracting the desired answer through statistical analysis of the measurement results. Cleverly choosing which measurements to perform may substantially reduce the number of total measurements needed to extract the same information at the same confidence level.

In most implementations of VQE, the expectation value of the Hamiltonian is extracted by evaluating the expectation value of Pauli strings (tensor products of Pauli operators) and later summing the results together. Since it is possible to simultaneously measure commuting observables, it is a commonly pursued strategy to find ways to group Pauli strings into sets in which all members commute with each other \cite{verteletskyi2020measurement, yen2020measuring, izmaylov2019unitary, crawford2021efficient}, although other approaches involving entangled measurements also exist \cite{hamamura2020efficient}.

Grouping the Pauli strings is done classically, but to actually perform the measurements, additional instructions for the quantum computer need to be provided. This involves generating a quantum circuit segment for each group of Pauli operators, which is responsible for performing a transformation equivalent to changing the measurement basis. The size and composition of this circuit depends both on the strategy and on the particular Pauli strings found in the given group.

These methods do not make quantum computations any easier in terms of qubit number or circuit depth. Instead, they simply reduce the amount of times a quantum computer needs to be used for the same task. This is beneficial because it reduces the total amount of time needed to complete a quantum computation.

\section{Theory and methods}
\begin{figure}[H]
\includegraphics[width=\linewidth]{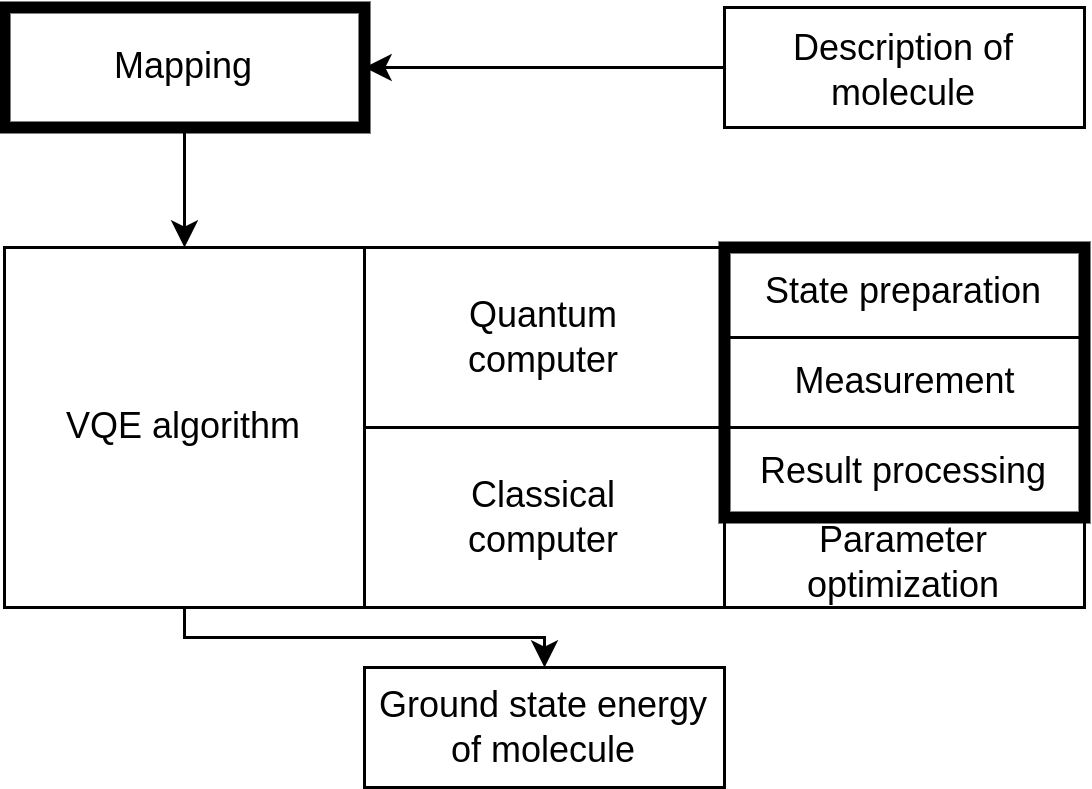}
\caption{Visual representation of information flow from input data, processing/calculation steps throughout hardware devices and to final end result. Steps covered in this research work are highlighted with bold margins.}
\label{fig:scope}
\end{figure}

In this section, we introduce a technique for achieving improvements in all three aspects discussed in the introduction. The technique comprises of a mapping method and a measurement method. The mapping method describes how to formulate problem constraints in an electronic structure problem and use them to map both the relevant electronic subspace and the Hamiltonian to the Hilbert space of a quantum computer. Since only the relevant subspace is mapped, this mapping simultaneously enforces constraints and reduces the number of necessary qubits. The measurement method describes how to measure the expectation value of the mapped Hamiltonian on a quantum computer and generates the necessary quantum circuits for this procedure. The measurement method can also be used independently of the mapping method, and when paired with conventional mappings such as Jordan-Wigner, is able to substantially reduce the number of different measurements needed to extract the expectation value of the Hamiltonian. Figure \ref{fig:scope} presents an overview of the computational steps that this work concerns.

Both methods are preprocessing steps for the VQE algorithm. They are executed classically and only need to be run once, regardless of the amount of iterations in the VQE algorithm. These methods do not put any limitations on which Ansatz can be used, but also do not offer a straightforward way to map circuits which were designed for the original, higher-qubit problem. Approaches which do not assume the number of qubits such as the hardware-efficient Ansatz \cite{kandala2017hardware} or qubit-ADAPT-VQE \cite{tang2021qubit} are unaffected by this issue.

\subsection{Mapping the states}
Suppose we are given a set of $K$ constraints, which the minimal energy state sought by the VQE algorithm ought to satisfy. Let us name the set of states that satisfy all $K$ constraints as \textit{valid states}. Each of these constraints is related to a physical feature, described by an operator $\hat{C}_k$ on the fermionic Fock space $\mathcal{F}$ with dimension $N$, where $k\in\{1,\hdots,K\}$. In principle, it should be possible to assign an operator to any physical quantity and explicit formulas can be found in the literature for electron number, electron number in a specific spin sector, spin multiplicity, spin projection onto an axis \cite{helgaker2014molecular}. In addition, spatial symmetries can be assigned operators, for example, by expressing them as a suitable combination of permutations \cite{setia2020reducing}.

In this language, if a constraint $k$ rules that a certain physical quantity should have a specific value, then valid states are eigenstates of the operator $\hat{C}_k$ that have a particular eigenvalue $\lambda_k$. This is equivalent to the statement that a valid state $\ps$ resides in the right null space $\mathcal{N}_k\subseteq\mathcal{F}$ of the operator $\hat{S}_k\equiv \hat{C}_k-\lambda_k\hat{I}$, where $\hat{I}$ is the identity operator on $\mathcal{F}$. This is because
\[
\text{If }&\hat{C}_k\ps=\lambda_k\ps\nonumber\\
\text{then }&\hat{S}_k\ps=\hat{C}_k\ps-\lambda_k\hat{I}\ps=0\nonumber\\
\text{so }&\ps\in\mathcal{N}_k\text{ by definition of }\mathcal{N}_k\label{q1}
\]

We could also consider a more general kind of constraint, which enumerates a list of $L$ permissible values. In this case, valid states are eigenstates of the operator $\hat{C}_k$ which have an eigenvalue that is contained within a list of $L\leq N$ allowed eigenvalues $\{\lambda_k^\ell\}$ where $\ell\in(1,\hdots,L)$. Valid states would then reside in the space
\[
\mathcal{N}_k=\bigcup_\ell^L\mathcal{N}_k^{\ell}
\label{w0}
\]
where $\mathcal{N}_k^\ell$ are the null spaces of the operators $\hat{S}_k^\ell\equiv \hat{C}_k-\lambda^\ell_k\hat{I}$, which are individually defined in the same manner as described for the single eigenvalue case.

To satisfy all of the different constraints at once, valid states must reside in the intersection $\mathcal{N}$ of all of the individual spaces $\mathcal{N}_k$.
\[
\mathcal{N}=\bigcap_k^K\mathcal{N}_k
\label{w1}
\]
Let the dimension of the subspace $\mathcal{N}$ be $M$. Since $\mathcal{N}$ is a subspace of $\mathcal{F}$, we know that $M\leq N$, but the exact value of $M$ will depend on which constraints are chosen. We can choose $M$ orthonormal states in $\mathcal{F}$, which span $\mathcal{N}$. We will use the notation $\ket{m}$ for these states, where $m\in(1,\hdots,M)$. Any state in the subspace $\ket{\psi_\mathcal{N}}\in\mathcal{N}$ can be expressed as a linear combination of these orthonormal states as
\[
\ket{\psi_\mathcal{N}}=\sum_m\alpha_m\ket{m}
\label{q2}
\]
where $\alpha_m$ are complex-valued coefficients.

\begin{figure}[H]
\includegraphics[width=\linewidth]{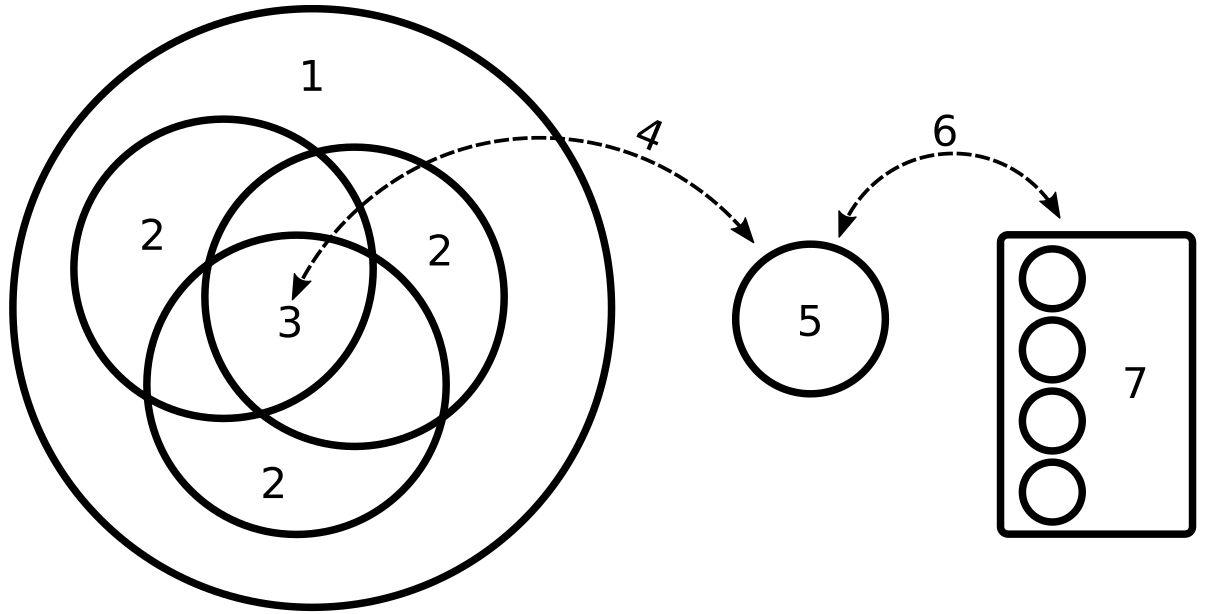}
\caption{Reduction of the original space size of the
problem (1; $\mathcal{F}$) by identifying the subspaces of valid states
defined by the problem constraints (2; $\mathcal{N}_k$). The intersection
of these subspaces (3; $\mathcal{N}$) is mapped (4; $\hat{D}$) to a newly
defined Hilbert space (5; $\mathcal{H}$), which can then be
represented (6; $\ket{m_*}$) on a quantum computer (7).}
\label{fig:states} 
\end{figure}

We will now construct a way to map all the states in $\mathcal{N}$ (and no others) to a quantum computer. For this, we define a linear operator $\hat{D}$, that maps states from $\mathcal{F}$ onto a new, $M$-dimensional Hilbert space $\mathcal{H}$. The operator is defined as
\[
\hat{D}\equiv\sum_m\ket{m_*}\bra{m}
\label{q3}
\]
where $\{\ket{m_*}\}$ is a complete orthonormal basis of $\mathcal{H}$, that is
\[
\sum_m\ket{m_*}\bra{m_*}=\hat{I}_\mathcal{H}
\label{q4}
\]
where $\hat{I}_\mathcal{H}$ is the identity operator on $\mathcal{H}$. We can now use definitions (\ref{q2}) and (\ref{q3}) to map any state in $\mathcal{N}$ to a new state in $\mathcal{H}$:
\[
\ket{\psi_\mathcal{H}}&\equiv\hat{D}\ket{\psi_\mathcal{N}}=\sum_{m,m'}\ket{m_*'}\braket{m'|\alpha_m|m}\nonumber\\
&=\sum_{m,m'}\delta_{mm'}\alpha_m\ket{m_*'}=\sum_m\alpha_m\ket{m_*}
\label{w2}
\]
After mapping, we can represent these states on a quantum computer by choosing unique computational basis states to represent each $\ket{m_*}$. Figure \ref{fig:states} provides a visual summary of the entire mapping procedure. The number of qubits required to have $M$ unique computational basis states is $\lceil\log_2M\rceil$.
This mapping is bijective between $\mathcal{N}$ and $\mathcal{H}$ because $\ket{\psi_\mathcal{N}}$ can be retrieved from $\ket{\psi_\mathcal{H}}$ by applying $\hat{D}^\dagger$:
\[
\hat{D}^\dagger\ket{\psi_\mathcal{H}}&=\sum_{m,m'}\ket{m'}\braket{m_*'|\alpha_m|m_*}=\sum_{m,m'}\delta_{mm'}\alpha_m\ket{m'}\nonumber\\
&=\sum_m\alpha_m\ket{m}=\ket{\psi_\mathcal{N}}
\label{ddag}
\]
Note that $\hat{D}$ can be applied to any state in $\mathcal{F}$, but when $\hat{D}^\dagger$ is subsequently applied, the original state is projected onto $\mathcal{N}$, which for states already in $\mathcal{N}$ is an identity operation. We can also see this by writing out the expression
\[
D^\dagger D&=\sum_{m,m'}\ket{m}\braket{m_*|m_*'}\bra{m'}=\sum_{m,m'}\ket{m}\delta_{mm'}\bra{m'}\nonumber\\
&=\sum_m\ket{m}\bra{m}
\label{dd}
\]
and noticing that it coincides with the definition of a projection operator onto the space spanned by $\{\ket{m}\}$, which is $\mathcal{N}$.

In summary, we have defined a bijective mapping using the operator $\hat{D}$, which constructs a representation of valid states on a quantum computer. Performing an unconstrained search in $\mathcal{H}$ with an algorithm such as VQE is equivalent to performing a search among valid states. The result of the search can always be retrieved using the inverse mapping shown in equation (\ref{ddag}).

\subsection{Mapping the Hamiltonian}
Let us define a complete orthonormal basis set $\{\ket{n}\}$ on $\mathcal{F}$, such that the first $M$ states are $\{\ket{m}\}$ (which span $\mathcal{N}$) and the rest are denoted as $\{\ket{p}\}$ (which span $\mathcal{F}-\mathcal{N}$), where $p\in(M+1,\hdots,N)$:
\[
\ket{n}=\left\{\begin{array}{lll}\ket{m}&\text{if}&1\leq n\leq M\\\ket{p}&\text{if}&M<n\leq N\end{array}\right.
\label{w3}
\]
An arbitrary state $\ket{a}\in\mathcal{F}$ can be written as
\[
\label{amp}
\ket{a}=\sum_n\ket{n}\braket{n|a}=\sum_m\ket{m}\braket{m|a}+\sum_p\ket{p}\braket{p|a}
\]
The Hamiltonian operator can be expressed as
\[
\label{ham}
\hat{H}&=\sum_{a,a'}h_{aa'}\ket{a}\bra{a'}
\]
where $h_{aa'}$ are complex-valued coefficients, such that $h_{aa'}$ and $h_{a'a}$ are complex conjugates of each other.
\[
h_{aa'}=h_{a'a}^*
\label{w4}
\]
This property follows from the requirement that the Hamiltonian is a Hermitian operator. The sets of states denoted by $a$ and $a'$ need not form an orthonormal or complete basis of any particular space, they only need to be sufficiently diverse that they are able to express all terms that the Hamiltonian contains. In the special case of real-valued coefficients, we have $h_{aa'}=h_{a'a}$. We make use of this property later in the measurement section (this will be pointed out), so part of the method is contingent on making an appropriate choice of $a$ and $a'$. Serendipitously, there is a straightforward way to do this. Open source quantum software development tools such as OpenFermion \cite{mcclean2020openfermion}, in conjunction with open source quantum chemistry software such as Psi4 \cite{turney2012psi4} express the Hamiltonian as a sum of terms, each of which consist of a product of a real-valued coefficient and various fermionic creation and annihilation operators. These operators are real-valued matrices in the occupation number basis, so if the occupation number basis is chosen to enumerate $a$ and $a'$, then all coefficients of the Hamiltonian are also real-valued. We can insert expression (\ref{amp}) into (\ref{ham}) to obtain
\[
\hat{H}&=\sum_{a,a'}h_{aa'}\left(\sum_{m}\ket{m}\braket{m|a}+\sum_{p}\ket{p}\braket{p|a}\right)\nonumber\\
&\times\left(\sum_{m'}\braket{a'|m'}\bra{m'}+\sum_{p'}\braket{a'|p'}\bra{p'}\right)
\label{w5}
\]
If the VQE algorithm performs a search among states the states $\ket{\psi_\mathcal{N}}\in\mathcal{N}$, then we can use the properties of those states
\[
\braket{\psi_\mathcal{N}|p}=\braket{p|\psi_\mathcal{N}}=\braket{\psi_\mathcal{N}|p'}=\braket{p'|\psi_\mathcal{N}}=0
\label{w6}
\]
to simplify the calculation of the expectation value of the Hamiltonian $\braket{\hat{H}}=\braket{\psi_\mathcal{N}|\hat{H}|\psi_\mathcal{N}}$. 
\[
\braket{\hat{H}}&=\sum_{a,a'}h_{aa'}\left(\sum_{m}\braket{\psi_\mathcal{N}|m}\braket{m|a}+\sum_{p}\braket{\psi_\mathcal{N}|p}\braket{p|a}\right)\nonumber\\
&\times\left(\sum_{m'}\braket{a'|m'}\braket{m'|\psi_\mathcal{N}}+\sum_{p'}\braket{a'|p'}\braket{p'|\psi_\mathcal{N}}\right)\nonumber\\
&=\sum_{a,a'}h_{aa'}\sum_{m}\braket{\psi_\mathcal{N}|m}\braket{m|a}\sum_{m'}\braket{a'|m'}\braket{m'|\psi_\mathcal{N}}\nonumber\\
&=\bra{\psi_\mathcal{N}}\sum_{m}\ket{m}\bra{m}\sum_{a,a'}h_{aa'}\ket{a}\bra{a'}\sum_{m'}\ket{m'}\bra{m'}\ket{\psi_\mathcal{N}}\nonumber\\
&=\braket{\psi_\mathcal{N}|\hat{D}^\dagger\hat{D}\hat{H}\hat{D}^\dagger\hat{D}|\psi_\mathcal{N}}=\braket{\psi_\mathcal{H}|\hat{H}_\mathcal{H}|\psi_\mathcal{H}}
\label{w7}
\]
where the transformed Hamiltonian $\hat{H}_\mathcal{H}$ has been defined as
\[
\hat{H}_\mathcal{H}\equiv\hat{D}\hat{H}\hat{D}^\dagger
\label{mappedh}
\]
which we can explicitly calculate by inserting expressions (\ref{q3}) and (\ref{ham}) into (\ref{mappedh}):
\[
\hat{H}_\mathcal{H}&=\sum_{m}\ket{m_*}\bra{m}\sum_{a,a'}h_{aa'}\ket{a}\bra{a'}\sum_{m'}\ket{m'}\bra{m_*'}\nonumber\\
&=\sum_{m,m'}h_{mm'}\ket{m_*}\bra{m_*'}
\label{newham}
\]
where
\[
h_{mm'}\equiv\sum_{aa'}h_{aa'}\braket{m|a}\braket{a'|m'}
\label{hdef}
\]
We are now, in principle, able to evaluate the expectation value of the Hamiltonian $\hat{H}$ in a valid state $\ket{\psi_\mathcal{N}}$ on the QC by instead evaluating the expectation value of the transformed Hamiltonian in the mapped state $\ket{\psi_\mathcal{H}}$:
\[
\braket{\hat{H}}=\sum_{mm'}h_{mm'}\braket{\psi_\mathcal{H}|m_*}\braket{m_*'|\psi_\mathcal{H}}
\label{w8}
\]
This requires fewer (or the same amount of) qubits than the original problem because the dimension of $\mathcal{H}$ is lower than (or equal to) the dimension of $\mathcal{F}$.

\subsection{Measuring the expectation value}
To evaluate the expectation value $\braket{\hat{H}}$, we will divide the terms in Equation (\ref{w8}) into two groups: those terms in which $m=m'$ and those where $m\neq m'$:
\[
\braket{\hat{H}}=\sum_mh_{mm}\braket{\psi_\mathcal{H}|m_*}\braket{m_*|\psi_\mathcal{H}}+\frac{1}{2}\sum_{m\neq m'}E_{mm'}
\label{twoparts}
\]
where
\[
E_{mm'}&\equiv h_{mm'}\braket{\psi_\mathcal{H}|m_*}\braket{m_*'|\psi_\mathcal{H}}\nonumber\\
&+h_{m'm}\braket{\psi_\mathcal{H}|m_*'}\braket{m_*|\psi_\mathcal{H}}
\label{emm0}
\]
Obtaining the terms in the sum where $m=m'$ is straightforward and can be calculated simultaneously, based on the same set of measurements. We can do this by following these steps:

1) Prepare the state $\ket{\psi_\mathcal{H}}$ on the QC

2) Perform a measurement on all qubits, which collapses the state $\ket{\psi_\mathcal{H}}$ to one of the computational basis states $\ket{m_*}$.

3) Repeat the above steps $n$ times and record which state the measurement collapsed to each time. From these results, extract the occurrence likelihood $p_m$ of each computational basis state via frequentist inference by dividing the number of occurrences $n_m$ with the total number of measurements $n$:
\[
p_m=\lim_{n\rightarrow\infty}\frac{n_m}{n}
\label{w9}
\]
On average, increasing $n$ improves the precision of (\ref{w9}).

4) Calculate the values of the terms in the Hamiltonian using Born's rule:
\[
\braket{\psi_\mathcal{H}|m_*}\braket{m_*|\psi_\mathcal{H}}=p_m
\label{w10}
\]
To calculate the terms where $m\neq m'$, we need to perform a different set of measurements. To do this, we will define a new unitary operator $\hat{R}$:
\[
\hat{R}^\dagger \hat{R}&=\hat{I}_\mathcal{H}
\label{runi}
\]
which we can insert into (\ref{emm0}) to obtain
\[
E_{mm'}&=h_{mm'}\braket{\psi_\mathcal{H}|\hat{R}^\dagger \hat{R}|m_*}\braket{m_*'|\hat{R}^\dagger \hat{R}|\psi_\mathcal{H}}\nonumber\\&+h_{m'm}\braket{\psi_\mathcal{H}|\hat{R}^\dagger\hat{R}|m_*'}\braket{m_*|\hat{R}^\dagger \hat{R}|\psi_\mathcal{H}}\nonumber\\
&=h_{mm'}\braket{\psi_R|\hat{R}|m_*}\braket{m_*'|\hat{R}^\dagger|\psi_R}\nonumber\\&+h_{m'm}\braket{\psi_R|\hat{R}|m_*'}\braket{m_*|\hat{R}^\dagger|\psi_R}
\label{r0}
\]
Where 
\[
\ket{\psi_R}\equiv \hat{R}\ket{\psi_\mathcal{H}}
\label{w11}
\]
Suppose $\hat{R}$ has the following properties:
\[
\label{r1}\hat{R}\ket{m_*}&=\frac{1}{\sqrt{2}}\left(\ket{m_*}+\ket{m_*'}\right)\\
\label{r2}\hat{R}\ket{m_*'}&=\frac{1}{\sqrt{2}}\left(\ket{m_*}-\ket{m_*'}\right)
\]
The factor $1/\sqrt{2}$ ensures that $\hat{R}$ is unitary. It is important to notice that these properties differentiate between the primed and unprimed states, which have up until now been treated on an equal footing. This introduces an ambiguity that needs to be resolved later: for every pair of states, there needs to exist a rule that determines which of those states will be denoted as primed. Inserting (\ref{r1}) and (\ref{r2}) into (\ref{r0}) yields
\[
E_{mm'}&=\frac{h_{mm'}}{2}\left(\braket{\psi_R|m_*}+\braket{\psi_R|m_*'}\right)\left(\braket{m_*|\psi_R}-\braket{m_*'|\psi_R}\right)\nonumber\\
&+\frac{h_{m'm}}{2}\left(\braket{\psi_R|m_*}-\braket{\psi_R|m_*'}\right)\left(\braket{m_*|\psi_R}+\braket{m_*'|\psi_R}\right)\nonumber\\
&=\frac{h_{mm'}+h_{m'm}}{2}\nonumber\\
&\times\braket{\psi_R|m_*}\braket{m_*|\psi_R}-\braket{\psi_R|m_*'}\braket{m_*'|\psi_R})\nonumber\\
&+\frac{h_{mm'}-h_{m'm}}{2}\nonumber\\
&\times(\braket{\psi_R|m_*'}\braket{m_*|\psi_R}-\braket{\psi_R|m_*}\braket{m_*'|\psi_R})\label{emm}
\]
We will now make a simplifying assumption that the Hamiltonian has only real-valued coefficients, which means that
\[
h_{mm'}=h_{m'm}
\label{realh}
\]
because $\hat{H}$ is Hermitian. This assumption can always be fulfilled through equations (\ref{ham}) and (\ref{hdef}), by choosing a suitable basis for the Hamiltonian. Using (\ref{realh}) reduces (\ref{emm}) to
\[
\label{res}E_{mm'}=h_{mm'}(\braket{\psi_R|m_*}\braket{m_*|\psi_R}-\braket{\psi_R|m_*'}\braket{m_*'|\psi_R})
\]
These terms are similar to the ones for the $m=m'$ case, and can be evaluated using the same four steps described earlier. The only difference is in the first step, where instead of $\ket{\psi_\mathcal{H}}$, we must prepare $\ket{\psi_R}$ instead. Preparing $\ket{\psi_R}$ entails preparing $\ket{\psi_\mathcal{H}}$ and then subsequently appending a quantum circuit which implements the $\hat{R}$ operator.
\subsection{Circuits for measurement}
A valid circuit for applying $\hat{R}$ is any circuit that is simultaneously consistent with equations (\ref{r1}) and (\ref{r2}). This can be achieved in many different ways and it is also possible to construct circuits which allow the measurement of multiple pairs of $m$ and $m'$ at the same time. Here we will provide a method to generate all necessary circuits for every pair, such that the number of unique circuits is no more than $2M$.

We will design a circuit that creates a superposition of two states which are fully anti-correlated in the qubits in which $\ket{m_*}$ and $\ket{m_*'}$ differ and fully correlated in the qubits in which they do not. This can be achieved with the following steps:

Step 1) For all terms appearing in the second half of (\ref{twoparts}), examine the representations of the states $\ket{m_*}$ and $\ket{m_*'}$ and determine in which qubits they differ. For example, by examining the term $\ket{0100}\bra{1101}$, we would say that they differ in the first and the last qubit. For ease of reference, let us name these qubits the \textit{active qubits}.

Step 2) Partition the terms into groups, where all terms within a group have the same set of active qubits. For example, the terms $\ket{0100}\bra{1101}$ and $\ket{0000}\bra{1001}$ belong to the same group. All terms within a single group shall be evaluated simultaneously, using the same circuit.

Step 3) For each group, choose one of the active qubits to be the \textit{control qubit}. The control qubit will be used to perform controlled two-qubit operations, so there may exist a preferential choice for this dictated by the quantum hardware.

Step 4) For every term, resolve the ambiguity introduced by (\ref{r1}) and (\ref{r2}) earlier. The state with $\ket{1}$ on the control qubit shall be the one denoted with a prime. This rule, in conjunction with the circuit described in the next step, is consistent with equations (\ref{r1}) and (\ref{r2}). In principle, the opposite choice would work equally well, if the right sides of (\ref{r1}) and (\ref{r2}) were switched, or if the circuit were suitably modified.

Step 5) Construct the circuit by placing a Hadamard gate on the control qubit and CNOT gates on either side, such that each other active qubit is targeted by either the control qubit or any other qubit that has already been targeted closer to the Hadamard.

\begin{figure*}
\includegraphics[width=0.8\linewidth]{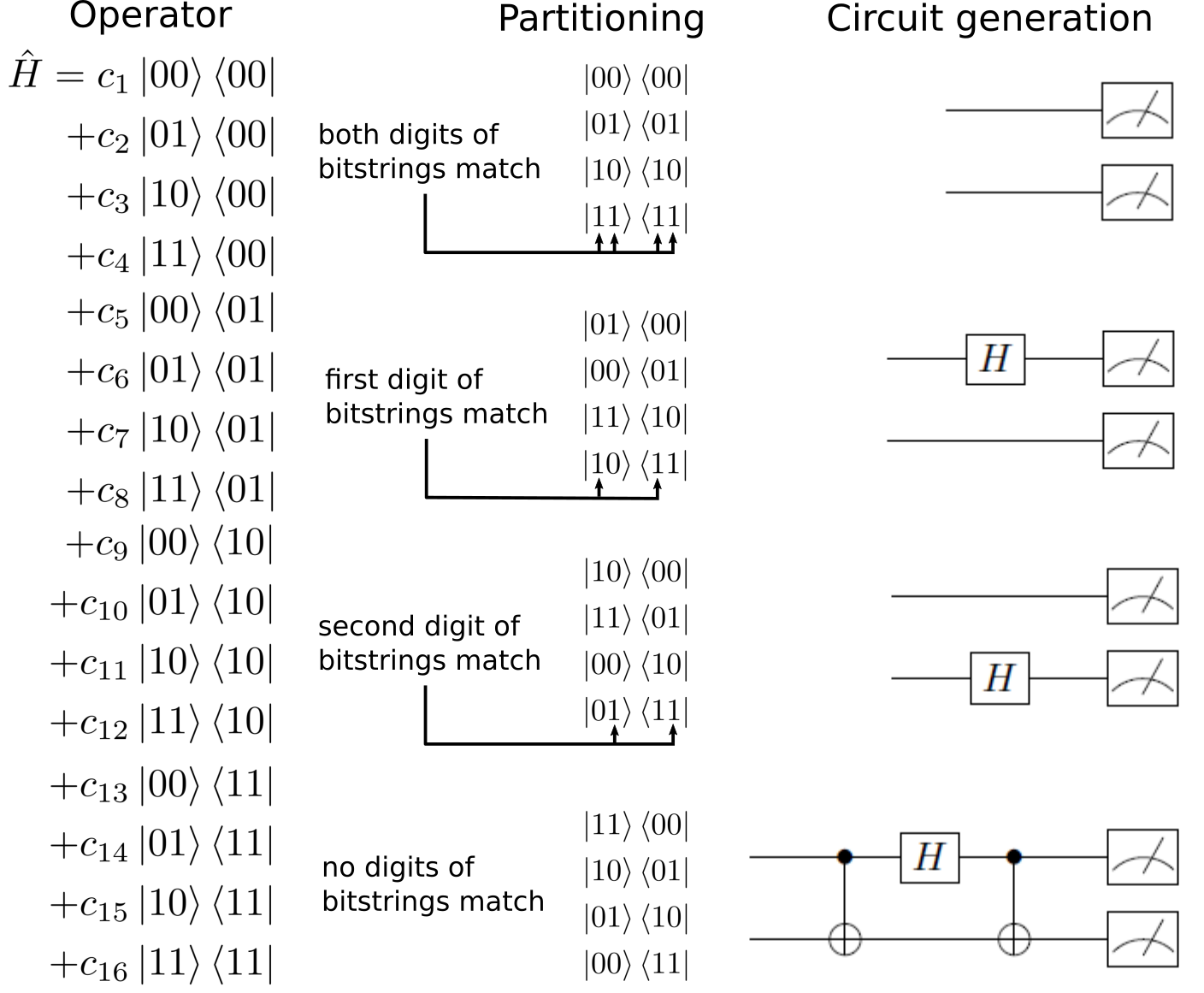}
\caption{\label{fig:circuitgen} The circuit generation procedure for a two-qubit Hamiltonian. Terms are grouped in to mutually measurable sets by comparing bitstrings. A circuit is generated for each group, such that the circuit implements a suitable $\hat{R}$ operator for all pairs in the group.}
\end{figure*}

An example of the whole procedure for a two qubit problem is provided in Figure \ref{fig:circuitgen}.

For more qubits, there are multiple equivalent circuits that can be generated, based on the description given in Step 5. Below are three examples in the case of four active qubits, with the control qubit marked with an arrow:
\[
\Qcircuit @C=1em @R=.7em {
&\lstick{\rightarrow}&\qw&\qw     &\qw     &\ctrl{1}&\gate{H}&\ctrl{1}&\qw     &\qw     &\qw\\
&                                        &\qw&\qw     &\ctrl{1}&\targ   &\qw     &\targ   &\ctrl{1}&\qw     &\qw\\
&                                        &\qw&\ctrl{1}&\targ   &\qw     &\qw     &\qw     &\targ   &\ctrl{1}&\qw\\
&                                        &\qw&\targ   &\qw     &\qw     &\qw     &\qw     &\qw     &\targ   &\qw
}
\label{circ1}
\]
\[
\Qcircuit @C=1em @R=.7em {
&\lstick{\rightarrow}&\qw&\ctrl{3}&\ctrl{2}&\ctrl{1}&\gate{H}&\ctrl{1}&\ctrl{2}&\ctrl{3}&\qw\\
&                                        &\qw&\qw     &\qw     &\targ   &\qw     &\targ   &\qw     &\qw     &\qw\\
&                                        &\qw&\qw     &\targ   &\qw     &\qw     &\qw     &\targ   &\qw     &\qw\\
&                                        &\qw&\targ   &\qw     &\qw     &\qw     &\qw     &\qw     &\targ   &\qw
}
\label{circ2}
\]
\[
\Qcircuit @C=1em @R=.7em {
&                                        &\qw&\targ    &\qw     &\qw     &\qw     &\targ    &\qw\\
&\lstick{\rightarrow}&\qw&\ctrl{-1}&\ctrl{1}&\gate{H}&\ctrl{2}&\ctrl{-1}&\qw\\
&                                        &\qw&\ctrl{1} &\targ   &\qw     &\qw     &\targ    &\qw\\
&                                        &\qw&\targ    &\qw     &\qw     &\targ   &\ctrl{-1}&\qw
}
\label{circ3}
\]
These circuits are equivalent, meaning that they implement the same unitary operation. A proof of this, using ZX calculus is provided in Appendix A. Flexibility in choosing between these circuits is especially beneficial for hardware architectures where not all qubits are connected to each other, because choosing the circuit based on the connectivity of the device avoids introducing additional swap operations into the circuit \cite{o2019generalized}.

To demonstrate that these example circuits are indeed suitable, we must confirm that equations (\ref{r1}) and (\ref{r2}) are satisfied. Regardless of the number of active qubits, this can be most easily seen from circuits which resemble the second circuit in the previous list of three examples, with the control qubit targeting all of the active qubits with CNOT gates. Such a circuit can be written in terms of operators as
\[
\hat{R}=\bigotimes_{i=1}^{n}\hat{C}^\text{ctrl}_{i}\otimes \hat{H}_\text{ctrl}\otimes\bigotimes_{i=1}^{n}\hat{C}^\text{ctrl}_{i}
\label{secondcircuit}
\]
where $\hat{H}_\text{ctrl}$ is the Hadamard gate applied to the control qubit, $\hat{C}^\text{ctrl}_i$ denotes a CNOT gate that targets the $i$-th active qubit and $n$ is the total number of active qubits (control qubit excluded). For the purposes of this proof, let us denote the states as $\ket{m_*}\leftrightarrow\ket{0xy}$ and $\ket{m_*'}\leftrightarrow\ket{1\bar{x}y}$, where the first number is the state of the control qubit, $x$ represents the states of the other active qubits and $y$ represents the states of the remaining qubits. Inserting (\ref{secondcircuit}) into equations (\ref{r1}) and (\ref{r2}) shows that they are indeed satisfied:
\[
\hat{R}\ket{m_*}&=\bigotimes_{i=1}^{n}\hat{C}^\text{ctrl}_{i}\otimes \hat{H}_\text{ctrl}\otimes\bigotimes_{i=1}^{n}\hat{C}^\text{ctrl}_{i}\ket{0xy}\nonumber\\
&=\bigotimes_{i=1}^{n}\hat{C}^\text{ctrl}_{i}\otimes\hat{H}_\text{ctrl}\ket{0xy}\nonumber\\
&=\frac{1}{\sqrt{2}}\bigotimes_{i=1}^{n}\hat{C}^\text{ctrl}_{i}\left(\ket{0xy}+\ket{1xy}\right)\nonumber\\&=\frac{1}{\sqrt{2}}\left(\ket{0xy}+\ket{1\bar{x}y}\right)=\frac{1}{\sqrt{2}}\left(\ket{m_*}+\ket{m_*'}\right)\label{w12}
\]
\[
\hat{R}\ket{m_*'}&=\bigotimes_{i=1}^{n}\hat{C}^\text{ctrl}_{i}\otimes \hat{H}_\text{ctrl}\otimes\bigotimes_{i=1}^{n}\hat{C}^\text{ctrl}_{i}\ket{1\bar{x}y}\nonumber\\
&=\bigotimes_{i=1}^{n}\hat{C}^\text{ctrl}_{i}\otimes \hat{H}_\text{ctrl}\ket{1xy}\nonumber\\
&=\frac{1}{\sqrt{2}}\bigotimes_{i=1}^{n}\hat{C}^\text{ctrl}_{i}\left(\ket{0xy}-\ket{1xy}\right)\nonumber\\&=\frac{1}{\sqrt{2}}\left(\ket{0xy}-\ket{1\bar{x}y}\right)=\frac{1}{\sqrt{2}}\left(\ket{m_*}-\ket{m_*'}\right)\label{w13}
\]

Finally, from Step 2, we can calculate an upper limit for the number of circuits $n_\text{circuits}^\text{max}$ needed to measure the expectation value of the Hamiltonian on $\log_2M$ qubits. The number of required circuits is equal to the number of ways it is possible to choose an unordered set of active qubits from the set of all qubits, which can be calculated as:
\[
\label{comb}
n_\text{circuits}^\text{max}=\sum_{i=0}^{\lceil\log_2M\rceil}\left(\begin{array}{c}\lceil\log_2M\rceil\\i\end{array}\right)=2^{\lceil\log_2M\rceil}
\]
The result lies between
\[
M\leq n_\text{circuits}^\text{max}<2M
\]
where the equality is achieved when $M$ is a power of 2. Note that this result is lower than the maximum number $n_\text{pauli}^\text{max}$ of different terms produced when mapping the same Hamiltonian to products of Pauli operators
\[
\label{maxpauli}
n_\text{pauli}^\text{max}=4^{\lceil\log_2M\rceil}-1
\]
Here the identity term has been left out because its expectation value is known without the need for a separate measurement. This result lies between
\[
M^2-1\leq n_\text{pauli}^\text{max}<2M^2-1
\]
again with the equality being achieved when $M$ is an exact power of two.
\section{Results}
The mapping and measurement methods are usable independently, so we will examine their performances both separately and together. To actually solve an electronic structure problem using VQE using both, it is additionally necessary to specify an Ansatz circuit.

\subsection{Mapping}
\begin{figure}
\includegraphics[width=\linewidth]{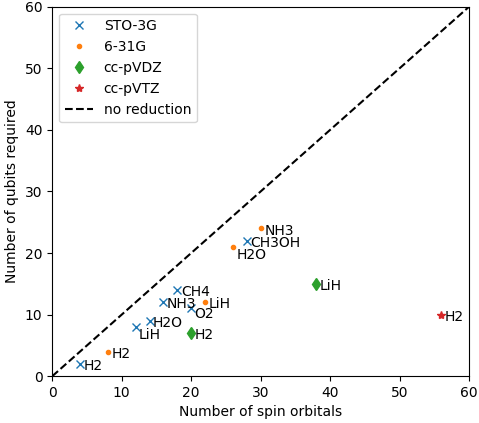}
\caption{\label{fig:qubitnumber}
The amount of qubits required for representing various molecules, with constraints on electron number and spin enforced with our mapping method. The different colours of crosses represent different atomic basis sets. The black dashed line is the requirement when using the Jordan-Wigner mapping.}
\end{figure}

One of the main motivations for using this mapping is to reduce the number of qubits required for the quantum simulation of molecules. Figure \ref{fig:qubitnumber} shows a selection of results for various molecules described in different basis sets. To obtain these results, we defined two constraints. One describes the number of electrons in the molecule, which we chose to be equal to the total charge of the nuclei, such that the molecule in question would be electrically neutral. The other constraint was on spin, where we specified that all molecules are in the singlet state, because that is what we expect the ground state to be for most of these molecules. The only exception is molecular oxygen ($\text{O}_2$), which we constrained to be in a triplet state. We present these results in Figure \ref{fig:qubitnumber}

In all cases, our mapping method reduced the number of required qubits needed to represent the molecules. As pointed out in Ref. \cite{chamaki2022compact}, for these particular constraints, it is possible to analytically calculate the number of required qubits $N_q$ as
\[
N_q=\left\lceil\log_2\left(\left(\begin{array}{c}M\\N^\uparrow\end{array}\right)\left(\begin{array}{c}M\\N^\downarrow\end{array}\right)\right)\right\rceil
\]
where $M$ is the number of spin orbitals in the basis set and $N_\uparrow$ and $N_\downarrow$ are the number of spin up and spin down electrons in the molecule, respectively. Our results agree with these analytical predictions for all cases, confirming that the mapping indeed does enforce these particular constraints correctly.

\begin{figure}
\includegraphics[width=\linewidth]{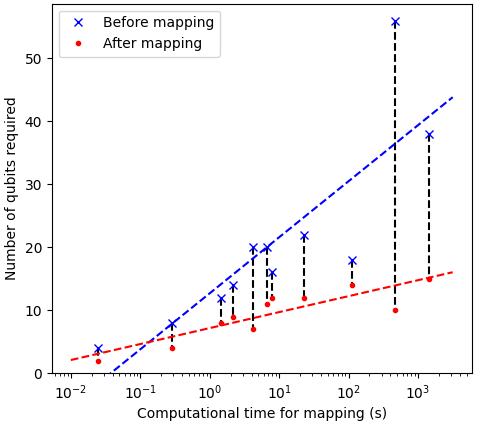}
\caption{\label{fig:maptime} Number of qubits required before and after the mapping, as a function of the time needed to perform the mapping. The black vertical dashed lines connect the pairs of points belonging to the same molecule. Blue and red dashed lines are obtained through linear regression of the likewise coloured points.}
\end{figure}

In Figure \ref{fig:maptime}, we show the performance and scaling of the computational demands of this mapping procedure for these constraints. All results are obtained using the same implementation on the same computer, using the molecules and basis sets presented in Figure \ref{fig:qubitnumber}. Only the three cases that had a qubit requirement over 20 were excluded from Figure \ref{fig:maptime} due to constraints on time and memory. The efficiency of the implementation could be improved upon and is left as a goal for future work. Since the horizontal axis in Figure \ref{fig:maptime} is logarithmic, the coloured dashed lines obtained through linear regression are equivalent to fitting an exponential function to the data as
\[
t=10^{a\cdot n+b}
\]
where $t$ is the computational time, $n$ is the qubit requirement and $a$ and $b$ are the slope and intercept of the line, respectively. From observing the regression lines, an exponential relationship between the final qubit requirement and the computational time spent is evident. The computational time also correlates with the number of orbitals and the number of qubits saved through the mapping, but characterizing this relationship as exponential is less accurate, due to the poorer fit of the regression line.

\begin{figure}
\includegraphics[width=\linewidth]{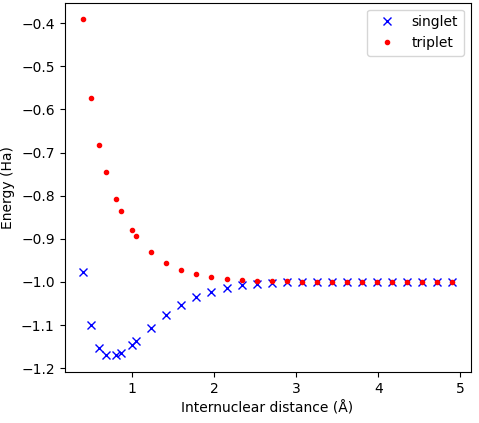}
\caption{\label{fig:h2curve} The energy of the singlet and triplet states of molecular hydrogen, as a function of the internuclear distance, using the complete set of 56 spin orbitals in the cc-pvtz basis. All points are calculated by classically extracting the minimal eigenvalue of the mapped Hamiltonian.}
\end{figure}

Figure \ref{fig:h2curve} shows the energy of the singlet and triplet states of hydrogen, obtained by performing the mapping with the corresponding constraints and retrieving the minimal eigenvalue classically, using the SciPy package \cite{2020SciPy-NMeth}. The full cc-pvtz basis set with 56 spin orbitals is used. The figure highlights three important results:

First, it shows that this mapping enables the use of more accurate basis sets for molecules. Most other mappings, even when combined with tapering, would require over 50 qubits to represent the full basis set on a quantum computer. Running a VQE algorithm of this size on quantum hardware has never been demonstrated, and a classical attempt to directly diagonalize a matrix with dimensions exceeding $2^{50}$ requires an exorbitant amount of computational resources. The problem was reduced to a 10-qubit problem for the singlet curve and a 9-qubit problem for the triplet curve, both easily solvable with a desktop computer.

Second, it allows us to confirm that the values obtained through this mapping are indeed correct, and no vital information is lost in the reduction of the problem dimension. In particular, the minimum value of the lower curve is -1.1723366673753 Hartree, which is equal to the CISD value provided by NIST \footnote{\url{https://cccbdb.nist.gov/energy3x.asp?method=15\&basis=6\&charge=0}} up to the number of digits they provide. Also, the values along the entire curve can be compared, for example, with the results in Ref \cite{hong2022accurate}, who perform an FCI calculation for this molecule with this basis set using Psi4 software \cite{turney2012psi4}

Third, we can see that by changing the constraints, it is possible to find the energies of excited states by finding only the lowest eigenvalue of the Hamiltonian. Since the lowest eigenvalue is the only eigenvalue that the VQE algorithm can find, then setting different constraints extends the applications of VQE, offering an alternative to other approaches for excited state VQE calcualtion such as Quantum Subspace Expansion \cite{mcclean2017hybrid, colless2018computation} or Variational Quantum Deflation \cite{higgott2019variational}

\subsection{Measurement}
To demonstrate how the measurement method works independently of the mapping method, we will first work with a widely known and easily reproducible example, which is the measurement of the ground-state energy of molecular hydrogen in the STO-3G basis set at an atomic distance of 0.75 \AA. The second quantized Hamiltonian can be obtained with quantum chemistry software such as Psi4 \cite{turney2012psi4}:
\newpage
\[\label{h2fh}
\hat{H}&=
0.70557
-1.24728\ a_1^\dagger a_1
-0.67284\ a_2^\dagger a_1^\dagger a_2 a_1\nonumber\\
&-0.18177\ a_2^\dagger a_1^\dagger a_4 a_3
-1.24728\ a_2^\dagger a_2\nonumber\\
&-0.48021\ a_3^\dagger a_1^\dagger a_3 a_1
-0.66198\ a_3^\dagger a_2^\dagger a_3 a_2\nonumber\\
&+0.18177\ a_3^\dagger a_2^\dagger a_4 a_1
-0.48127\ a_3^\dagger a_3\nonumber\\
&+0.18177\ a_4^\dagger a_1^\dagger a_3 a_2
-0.66198\ a_4^\dagger a_1^\dagger a_4 a_1\nonumber\\
&-0.48021\ a_4^\dagger a_2^\dagger a_4 a_2
-0.18177\ a_4^\dagger a_3^\dagger a_2 a_1\nonumber\\
&-0.69582\ a_4^\dagger a_3^\dagger a_4 a_3
-0.48127\ a_4^\dagger a_4
\]
where all the numbers are in units of Hartree and the indices on the creation and annihilation operators refer to spin indexes, with even numbers representing alpha orbitals and odd numbers representing beta orbitals. If this were to be transformed with the Jordan-Wigner transformation (Equations (\ref{jw1}) and (\ref{jw2})), it would become
\[\label{jwh}
\hat{H}_\text{JW}&=
-0.10973
-0.04544\ \hat{\sigma}^x_1\hat{\sigma}^x_2\hat{\sigma}^y_3\hat{\sigma}^y_4\nonumber\\
&+
0.04544\ \hat{\sigma}^x_1 \hat{\sigma}^y_2 \hat{\sigma}^y_3 \hat{\sigma}^x_4
+
0.04544\ \hat{\sigma}^y_1 \hat{\sigma}^x_2 \hat{\sigma}^x_3 \hat{\sigma}^y_4\nonumber\\
&-0.04544\ \hat{\sigma}^y_1 \hat{\sigma}^y_2 \hat{\sigma}^x_3 \hat{\sigma}^x_4
+0.16988\ \hat{\sigma}^z_1 \nonumber\\
&+
0.16821\ \hat{\sigma}^z_1 \hat{\sigma}^z_2
+0.12005\ \hat{\sigma}^z_1 \hat{\sigma}^z_3\nonumber\\
&+0.16549\ \hat{\sigma}^z_1 \hat{\sigma}^z_4 +
0.16988\ \hat{\sigma}^z_2\nonumber\\
&+0.16549\ \hat{\sigma}^z_2 \hat{\sigma}^z_3 +
0.12005\ \hat{\sigma}^z_2 \hat{\sigma}^z_4\nonumber\\
&-0.21886\ \hat{\sigma}^z_3 +
0.17395\ \hat{\sigma}^z_3 \hat{\sigma}^z_4
-0.21886\ \hat{\sigma}^z_4
\]
$\hat{H}_\text{JW}$ has a total of $14$ unique Pauli strings, each of which require repeated execution of a unique circuit to evaluate the expectation value, unless additional grouping methods are used. On the other hand, we can use the same basis states as the Pauli operators to express the Hamiltonian in the form of (\ref{ham}) as 
\[
\label{ex1h}
\hat{H}&=
0.90148\ \ket{0000}\bra{0000}
-0.45524\ \ket{0001}\bra{0001}\nonumber\\
&-0.45524\ \ket{0010}\bra{0010}
-1.11615\ \ket{0011}\bra{0011}\nonumber\\
&+0.18177\ \ket{1100}\bra{0011}
+0.33374\ \ket{0100}\bra{0100}\nonumber\\
&-0.54278\ \ket{0101}\bra{0101}
-0.36101\ \ket{0110}\bra{0110}\nonumber\\
&-0.18177\ \ket{1001}\bra{0110}
-0.54171\ \ket{0111}\bra{0111}\nonumber\\
&+0.33374\ \ket{1000}\bra{1000}
-0.18177\ \ket{0110}\bra{1001}\nonumber\\
&-0.36101\ \ket{1001}\bra{1001}
-0.54278\ \ket{1010}\bra{1010}\nonumber\\
&-0.54171\ \ket{1011}\bra{1011}
+0.18177\ \ket{0011}\bra{1100}\nonumber\\
&+0.43884\ \ket{1100}\bra{1100}
+0.22430\ \ket{1101}\bra{1101}\nonumber\\
&+0.22430\ \ket{1110}\bra{1110}
+0.70557\ \ket{1111}\bra{1111}
\]
Although this contains more terms than (\ref{jwh}), it can be evaluated by performing measurements on just two unique circuits, because in each individual term, the two basis states are either identical or differ in all qubits. For both circuits, we prepare the ground state using a well-known circuit \cite{o2016scalable}, not involving any parameter optimization at this point. For the terms with identical basis states, we just perform measurements in the computational basis, as shown in Circuit (\ref{circuit1}).

\begin{widetext}
\[\label{circuit1}
         \Qcircuit @C=1em @R=.7em {
&\ket{0}\quad&\gate{H}         &\ctrl{1}&\qw     &\qw     &\qw             &\qw     &\qw     &\ctrl{1}&\gate{H}          &\meter\\
&\ket{0}\quad&\gate{H}         &\targ   &\ctrl{1}&\qw     &\qw             &\qw     &\ctrl{1}&\targ   &\gate{H}          &\meter\\
&\ket{1}\quad&\gate{H}         &\qw	    &\targ   &\ctrl{1}&\qw 			   &\ctrl{1}&\targ   &\qw	  &\gate{H}          &\meter\\  
&\ket{1}\quad&\gate{R_x(\pi/2)}&\qw     &\qw 	 &\targ   &\gate{R_z(0.23)}&\targ   &\qw 	 &\qw     &\gate{R_x(-\pi/2)}&\meter
}
\]
\end{widetext}
Then, by extracting the probabilities with a state vector simulator in Pennylane \cite{bergholm2018pennylane}, and using Equation (\ref{w10}), we obtain
\[
\braket{\psi_g|0011}\braket{0011|\psi_g}&=0.98683\label{psig1}\\ \braket{\psi_g|1100}\braket{1100|\psi_g}&=0.01316\label{psig2}
\]
with all other probabilities being zero.
The remaining terms in the Hamiltonian (\ref{ex1h}) can be measured with Circuit (\ref{circuit2}), which includes an additional circuit segment before the measurement. That segment transforms the prepared state $\ket{\psi_g}$ into the state $\ket{\psi_R}$, which enables us to follow Equation (\ref{res}) and obtain Equations (\ref{firstterm}) and (\ref{secondterm})\\
\begin{widetext}
\[\label{circuit2}
         \Qcircuit @C=1em @R=.7em {
&\ket{0}\quad&\gate{H}         &\ctrl{1}&\qw     &\qw     &\qw             &\qw     &\qw     &\ctrl{1}&\gate{H}          &\qw     &\qw     &\ctrl{1}&\gate{H}&\ctrl{1}&\qw     &\qw     &\meter\\
&\ket{0}\quad&\gate{H}         &\targ   &\ctrl{1}&\qw     &\qw             &\qw     &\ctrl{1}&\targ   &\gate{H}          &\qw     &\ctrl{1}&\targ   &\qw     &\targ   &\ctrl{1}&\qw     &\meter\\
&\ket{1}\quad&\gate{H}         &\qw	    &\targ   &\ctrl{1}&\qw 			   &\ctrl{1}&\targ   &\qw	  &\gate{H}          &\ctrl{1}&\targ   &\qw     &\qw     &\qw     &\targ   &\ctrl{1}&\meter\\  
&\ket{1}\quad&\gate{R_x(\pi/2)}&\qw     &\qw 	 &\targ   &\gate{R_z(0.23)}&\targ   &\qw 	 &\qw     &\gate{R_x(-\pi/2)}&\targ   &\qw     &\qw     &\qw     &\qw     &\qw     &\targ   &\meter
}\]
\[
\braket{\psi_g|1001}\braket{0110|\psi_g}+\braket{\psi_g|0110}\braket{1001|\psi_g}&=\braket{\psi_R|0110}\braket{0110|\psi_R}-\braket{\psi_R|1001}\braket{1001|\psi_R}\label{firstterm}\\
\braket{\psi_g|1100}\braket{0011|\psi_g}+\braket{\psi_g|0011}\braket{1100|\psi_g}&=\braket{\psi_R|0011}\braket{0011|\psi_R}-\braket{\psi_R|1100}\braket{1100|\psi_R}\label{secondterm}
\]
\end{widetext}
Performing the measurements with this circuit yields
\[
\braket{\psi_R|0011}\braket{0011|\psi_R}&=0.38601\label{psir1}\\ \braket{\psi_R|1100}\braket{1100|\psi_R}&=0.61399\label{psir2}
\]
with all other probabilities being zero. Consequently, the right side of Equation (\ref{firstterm}) evaluates to zero. Inserting (\ref{psir1}) and (\ref{psir2}) into (\ref{secondterm}) and subsequently combining (\ref{ex1h}), (\ref{psig1}), (\ref{psig2}) and (\ref{secondterm}) we obtain the expectation value of the Hamiltonian
\[
&\braket{\psi_g|\hat{H}|\psi_g}=-1.11615\cdot0.98683+0.43884\cdot0.01316\nonumber\\
&+0.18177\cdot(0.38601-0.61399)=-1.13712
\]
which is precisely the value of the ground state energy of this Hamiltonian, which we expected to obtain when preparing the ground state $\ket{\psi_g}$.

Table \ref{tab:circuits} compares this method to other strategies \cite{verteletskyi2020measurement,yen2020measuring,izmaylov2019unitary,crawford2021efficient} for reducing the number of required measurements, showing a selection of molecules for which results have been published in the corresponding references. All entries in the table are obtained using the STO-3G basis set, using the Jordan-Wigner mapping for Refs \cite{verteletskyi2020measurement,yen2020measuring,izmaylov2019unitary} and the symmetry conserving Bravyi-Kitaev mapping for Ref \cite{crawford2021efficient}. Note that although the last mapping reduces the number of qubits by two, the number of terms in the Hamiltonian remains the same for all cases presented in the table besides $\text{H}_2$, so the comparison remains fair. In all cases, the method introduced in this work outperforms the naive approach and the approach of partitioning Pauli operators into qubit wise commuting sets \cite{verteletskyi2020measurement}. 
\begin{table}
\caption{\label{tab:circuits}
Comparison of the number of unique circuits needed to measure the expectation value of Hamiltonians describing various molecules. $N$ shows the number of qubits in those circuits and 'Naive' refers to number of circuits needed when all terms in the Hamiltonian are measured separately, excluding the identity term.}
\begin{ruledtabular}
\begin{tabular}{cccccccc}
&N&Naive&Ref \cite{verteletskyi2020measurement}&Ref \cite{yen2020measuring}&Ref \cite{izmaylov2019unitary}&Ref \cite{crawford2021efficient}&This work\\
\hline
$\text{H}_2$&4&14&3&2&11&2&2\\
$\text{BeH}_2$&14&665&203&28&110&-&94\\
$\text{H}_2$O&14&1085&322&43&127&51&162\\
$\text{NH}_3$&16&3608&1201&98&251&118&461\\
$\text{N}_2$&20&2950&1187&128&268&78&378\\
\end{tabular}
\end{ruledtabular}
\end{table}
\subsection{Combined mapping and measurement}
The mapping and measurement methods are designed to work with each other, because the output of the mapping method coincides with the input of the measurement method. Table \ref{tab:circuitsaftermap} shows the number of circuits needed to evaluate the mapped Hamiltonians of various molecules. All molecules in this table use the STO-3G basis set and the constraints used to define the mapping are the same ones as used earlier, on electron number and spin. Table \ref{tab:circuitsh2} provides the same information, but for a fixed molecule (hydrogen), while varying the bond distance and the number of orbitals mapped, which were chosen in such a way as to provide a direct comparison with Table I in Ref. \cite{chamaki2022compact}.

A general pattern that we have identified is that the number of circuits needed for each mapped Hamiltonian is close to the maximum number, calculated in Equation (\ref{comb}). Note that this was not true when the mapping method was omitted, as can be seen from Table \ref{tab:circuits}. With the mapped Hamiltonian, there is almost no deviation from this pattern, regardless of molecule, bond distance, and basis set size, as shown in Tables \ref{tab:circuitsaftermap} and \ref{tab:circuitsh2}. This may even increase the number of circuits needed if mapping is used, as exemplified by the $\text{H}_2$O molecule present in both Tables \ref{tab:circuits} and \ref{tab:circuitsaftermap}. Our interpretation of this result is that the mapping produces a denser Hamiltonian than the original, which is an expected side effect of encoding the same amount of relevant information into an object with fewer dimensions. In many cases, this creates a trade off between the number of qubits reduced and the number of measurements required, although in some cases, such as the rightmost columns in Table \ref{tab:circuitsh2}, improvements are achieved in both simultaneously.

This behaviour is also observed in Ref \cite{chamaki2022compact}, where they produce close to all possible Pauli strings, calculated in Equation (\ref{maxpauli}), which without additional Pauli grouping techniques would require quadratically more circuits than our method.
\begin{table}
\caption{\label{tab:circuitsaftermap}
Comparison of the number of unique circuits needed to measure the expectation value of Hamiltonians describing various molecules. $N$ shows the number of qubits in those circuits and 'Naive' refers to number of circuits needed when all terms in the Hamiltonian are measured separately, excluding the identity term.}
\begin{ruledtabular}
\begin{tabular}{ccccc}
&$N$ before&$N$ after&Terms before&Circuits after\\
\hline
$\text{H}_2$&4&2&14&2\\
LiH&12&8&630&255\\
$\text{H}_2$O&14&9&1085&512\\
Be$\text{H}_2$&14&11&665&1943\\
C$\text{H}_4$&18&14&9260&16384\\
\end{tabular}
\end{ruledtabular}
\end{table}
\begin{table*}
\caption{\label{tab:circuitsh2}
Number of circuits required to measure the expectation value of the mapped Hamiltonian of the Hyrdogen molecule. Numbers on the top of each column denote (the number of spin orbitals mapped)/(number of qubits required) and numbers in the table denote (the number of terms in the original Hamiltonian, excluding the identity term)/(the number of circuits required for the mapped Hamiltonian)}
\begin{ruledtabular}
\begin{tabular}{ccccccccc}
Bond distances&4/2&8/4&10/5&16/6&22/7&30/8&44/9&56/10\\
\hline
0.35\AA&14/2&184/15&251/30&1176/63&4380/128&14237/256&70094/512&191304/1024\\
0.55\AA&14/2&184/15&251/30&1236/63&4389/128&13425/256&68286/512&188088/1024\\
0.7\AA&14/2&184/15&251/30&1236/63&4389/128&13425/256&70110/512&191320/1024\\
1.4\AA&14/2&184/15&251/30&1236/63&4389/128&16169/256&70110/512&191312/1024\\
2.8\AA&14/2&184/15&443/31&1236/63&4389/128&12073/256&66158/512&191320/1024\\
4.4\AA&14/2&184/15&443/31&1236/63&4389/128&11753/256&64906/512&189932/1024\\
6.0\AA&14/2&184/15&443/31&1236/63&4389/128&13441/256&70130/512&204876/1024
\end{tabular}
\end{ruledtabular}
\end{table*}

To perform VQE on a quantum computer, an Ansatz needs to specified. Although the methods presented here do not force any restrictions on the choice of Ansatz, they do create some additional nuances. If the desired Ansatz is one that was specifically designed for the original problem before the mapping, an additional mapping step would have to be performed for the Ansatz. This is necessary from a purely compatibility standpoint, because the mapping will often remove qubits from the circuit. Although there is no mathematical restriction that prohibits doing this, we have not developed such a method yet and leave this as a subject for future work. However, some classes of Ansatz are easier to import than others, if they are formulated in a way which is agnostic to the number of qubits in the problem. For example, adaptively generated Ansatz circuits \cite{tang2021qubit} can be used with no modifications.
\section{Conclusion}
We have developed a method to improve the quantum simulation of the ground state energy of molecules by enforcing problem constraints while simultaneously reducing the amount of quantum resources needed. We have split this method into a mapping and a measurement part and shown that they can be used independently of each other. We have shown that with the mapping method, it is possible to achieve a significant reduction in the number of required qubits, without loss of information or accuracy contained within the original Hamiltonian. This reduction is especially pronounced when using increasingly larger basis sets. The scaling of the computational time needed to implement this method is best characterized as an exponential with respect to the number of qubits needed after the mapping. We have developed a method for measuring the expectation value of a Hamiltonian or any other Hermitian operator with real coefficients. This is done by generating circuits based on groups of computational basis state pairs rather than a Pauli matrix decomposition of the Hamiltonian. We find that when used separately from the mapping, this measurement method reduces the number of required circuits by a moderate amount, which we compared to other existing methods based on grouping Pauli operators. We find that when the mapping and measurement methods are used together, the number of required circuits is often close to the theoretical limit, which is no more than twice the dimension of the Hilbert space on which the Hamiltonian is defined.
\begin{acknowledgments}
We wish to acknowledge financial support from the PhotoQ project, funded by Innovation Fund Denmark. We would like to thank Artur Izmaylov for bringing reference \cite{chamaki2022compact}
 to our attention.
\end{acknowledgments}
\newpage
\appendix

\section{ZX calculus and circuit equivalence}
It is important to establish that circuits (\ref{circ1})-(\ref{circ3}) are equivalent to each other. A very general way to do this is by using ZX calculus \cite{van2020zx,coecke2021kindergarden}.

The rules of ZX calculus are detailed in Figure 1 of Ref \cite{van2020zx}, and the conventions used in this document are the same as in that document. For fast referencing, the rules used here are:
\begin{itemize}
\item
Spiders can be freely moved around, as long as their connections stay the same.
\item
Rule (f) is the spider fusion rule.
\item
Rule (b) is the bialgebra rule.
\item
In the paper, a decomposition of the Hadamard gate is derived in eq (45). We will refer to this decomposition as Rule (45) in this document.
\end{itemize}
First, we will derive a useful transformation:
\begin{widetext}
\[
\includegraphics[width=0.9\linewidth]{img/first.tikz}
\]
\end{widetext}
Transformation (A1) uses the following rules of ZX calculus from left to right:
\begin{enumerate}
    \item Rule (f) is used to merge the black spiders on the bottom.
    \item Rule (f) is used to split the black spiders on the bottom. The black spider on the top is moved for visual clarity.
    \item Rule (b) is applied to the four spiders in the center.
    \item The spiders are moved for visual clarity.
\end{enumerate}
\subsection{Two qubit circuits}
We will derive the ZX diagram to represent the following circuit:\\
\centerline{
         \Qcircuit @C=1em @R=.7em {
&\qw&\ctrl{1}&\gate{H}&\ctrl{1}&\qw\\
&\qw&\targ   &\qw     &\targ   &\qw
}}
Spiders with phases have the angle of the phase labeled directly above them.
\begin{widetext}
\[
\includegraphics[width=0.9\linewidth]{img/second.tikz}
\]
\end{widetext}

\begin{enumerate}
    \item The leftmost diagram is a direct translation of the circuit to a ZX-diagram.
    \item Rule (45) is applied.
    \item Rule (f) is used to move the white spiders through each other
    \item Transformation (1) is applied, in conjunction with rule (f) to merge the topmost white spiders.
\end{enumerate}

\subsection{Three qubit circuits}
We will derive the ZX diagram to represent the following circuit:\\
\centerline{
         \Qcircuit @C=1em @R=.7em {
&\qw     &\ctrl{2}&\ctrl{1}&\gate{H}&\ctrl{1}&\ctrl{2}&\qw\\
&\qw     &\qw     &\targ   &\qw     &\targ   &\qw     &\qw\\
&\qw     &\targ   &\qw     &\qw     &\qw     &\targ   &\qw
}}
\begin{widetext}
\[
\includegraphics[width=0.9\linewidth]{img/third.tikz}
\]
\end{widetext}

\begin{enumerate}
    \item The leftmost diagram is a direct translation of the circuit to a ZX-diagram.
    \item Transformation (A2) is applied
    \item Rule (f) is used to move the white spiders through each other
    \item Transformation (A1) is applied on the top and bottom line, in conjunction with rule (f) to merge the topmost white spiders.
\end{enumerate}
We will derive the ZX diagram to represent the following circuit:\\
\centerline{
         \Qcircuit @C=1em @R=.7em {
&\qw     &\qw     &\ctrl{1}&\gate{H}&\ctrl{1}&\qw     &\qw\\
&\qw     &\ctrl{1}&\targ   &\qw     &\targ   &\ctrl{1}&\qw\\
&\qw     &\targ   &\qw     &\qw     &\qw     &\targ   &\qw
}}
\begin{widetext}
\[
\includegraphics[width=0.7\linewidth]{img/fourth.tikz}
\]
\end{widetext}
\begin{enumerate}
    \item The leftmost diagram is a direct translation of the circuit to a ZX-diagram.
    \item Transformation (A2) is applied
    \item Transformation (A1) is applied on the bottom two lines, in conjunction with rule (f) to merge the topmost white spiders.
\end{enumerate}
\subsection{Circuits with more qubits}
Just as the diagram for the two qubit circuit was used in the derivation of the three qubit diagrams, the three qubit diagrams can be used in the derivation of four qubit diagrams and so on. Adding additional qubits targeted by CNOTs on either side result in the same ZX diagram (depicted below) regardless of which qubits are used as the control qubits (as long as the control qubits have been previously targeted).
\begin{center}
    \includegraphics[width=0.4\linewidth]{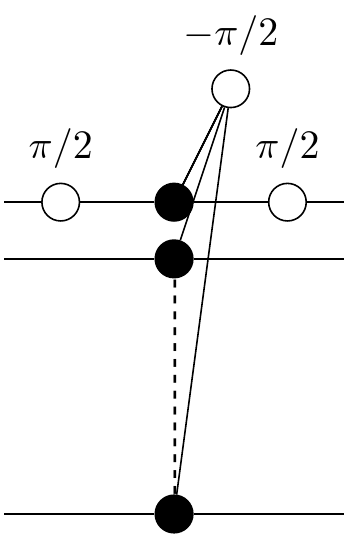}
\end{center}
Although not demonstrated in this document, this also applies to cases where the control qubits on the left and right side are different, for example a circuit like\\
\centerline{
         \Qcircuit @C=1em @R=.7em {
&\qw     &\ctrl{2}&\ctrl{1}&\gate{H}&\ctrl{1}&\qw     &\qw\\
&\qw     &\qw     &\targ   &\qw     &\targ   &\ctrl{1}&\qw\\
&\qw     &\targ   &\qw     &\qw     &\qw     &\targ   &\qw
}}

Since ZX calculus is sound and complete, then only circuits that represent the same linear map can be expressed with the same ZX diagram. That means that any circuit, whose representation as a ZX diagram can be transformed into the one here using the rewrite rules of ZX calculus, represents the same linear map as the example circuits presented here. This constitutes a sufficiency criteria: if a given circuit can be transformed into this diagram, then it may be used instead of the circuits shown here. This is relevant for NISQ devices, where certain circuits may be preferable to others. For example, Circuit (\ref{circ3}) is shallower than Circuits (\ref{circ1}) and (\ref{circ2}), meaning it is less prone to decoherence. Another example is choosing a circuit based on the connectivity of the device. If the physical device has no possibility of directly applying a CNOT gate between the bottom two qubits in this example, then Circuit (\ref{circ2}) can be used, as it does not contain such operations, whereas the other two do. Also, as long as the ZX diagram is realized, the circuit need not necessarily contain Hadamard or CNOT gates, if they can be substituted with different gates perhaps native to a particular hardware.

\bibliography{bib}

\begin{thebibliography}{55}%
\makeatletter
\providecommand \@ifxundefined [1]{%
 \@ifx{#1\undefined}
}%
\providecommand \@ifnum [1]{%
 \ifnum #1\expandafter \@firstoftwo
 \else \expandafter \@secondoftwo
 \fi
}%
\providecommand \@ifx [1]{%
 \ifx #1\expandafter \@firstoftwo
 \else \expandafter \@secondoftwo
 \fi
}%
\providecommand \natexlab [1]{#1}%
\providecommand \enquote  [1]{``#1''}%
\providecommand \bibnamefont  [1]{#1}%
\providecommand \bibfnamefont [1]{#1}%
\providecommand \citenamefont [1]{#1}%
\providecommand \href@noop [0]{\@secondoftwo}%
\providecommand \href [0]{\begingroup \@sanitize@url \@href}%
\providecommand \@href[1]{\@@startlink{#1}\@@href}%
\providecommand \@@href[1]{\endgroup#1\@@endlink}%
\providecommand \@sanitize@url [0]{\catcode `\\12\catcode `\$12\catcode
  `\&12\catcode `\#12\catcode `\^12\catcode `\_12\catcode `\%12\relax}%
\providecommand \@@startlink[1]{}%
\providecommand \@@endlink[0]{}%
\providecommand \url  [0]{\begingroup\@sanitize@url \@url }%
\providecommand \@url [1]{\endgroup\@href {#1}{\urlprefix }}%
\providecommand \urlprefix  [0]{URL }%
\providecommand \Eprint [0]{\href }%
\providecommand \doibase [0]{https://doi.org/}%
\providecommand \selectlanguage [0]{\@gobble}%
\providecommand \bibinfo  [0]{\@secondoftwo}%
\providecommand \bibfield  [0]{\@secondoftwo}%
\providecommand \translation [1]{[#1]}%
\providecommand \BibitemOpen [0]{}%
\providecommand \bibitemStop [0]{}%
\providecommand \bibitemNoStop [0]{.\EOS\space}%
\providecommand \EOS [0]{\spacefactor3000\relax}%
\providecommand \BibitemShut  [1]{\csname bibitem#1\endcsname}%
\let\auto@bib@innerbib\@empty
\bibitem [{\citenamefont {Peruzzo}\ \emph {et~al.}(2014)\citenamefont
  {Peruzzo}, \citenamefont {McClean}, \citenamefont {Shadbolt}, \citenamefont
  {Yung}, \citenamefont {Zhou}, \citenamefont {Love}, \citenamefont
  {Aspuru-Guzik},\ and\ \citenamefont {O’brien}}]{peruzzo2014variational}%
  \BibitemOpen
  \bibfield  {author} {\bibinfo {author} {\bibfnamefont {A.}~\bibnamefont
  {Peruzzo}}, \bibinfo {author} {\bibfnamefont {J.}~\bibnamefont {McClean}},
  \bibinfo {author} {\bibfnamefont {P.}~\bibnamefont {Shadbolt}}, \bibinfo
  {author} {\bibfnamefont {M.-H.}\ \bibnamefont {Yung}}, \bibinfo {author}
  {\bibfnamefont {X.-Q.}\ \bibnamefont {Zhou}}, \bibinfo {author}
  {\bibfnamefont {P.~J.}\ \bibnamefont {Love}}, \bibinfo {author}
  {\bibfnamefont {A.}~\bibnamefont {Aspuru-Guzik}},\ and\ \bibinfo {author}
  {\bibfnamefont {J.~L.}\ \bibnamefont {O’brien}},\ }\bibfield  {title}
  {\bibinfo {title} {A variational eigenvalue solver on a photonic quantum
  processor},\ }\href@noop {} {\bibfield  {journal} {\bibinfo  {journal}
  {Nature communications}\ }\textbf {\bibinfo {volume} {5}},\ \bibinfo {pages}
  {1} (\bibinfo {year} {2014})}\BibitemShut {NoStop}%
\bibitem [{\citenamefont {Cerezo}\ \emph {et~al.}(2021)\citenamefont {Cerezo},
  \citenamefont {Arrasmith}, \citenamefont {Babbush}, \citenamefont {Benjamin},
  \citenamefont {Endo}, \citenamefont {Fujii}, \citenamefont {McClean},
  \citenamefont {Mitarai}, \citenamefont {Yuan}, \citenamefont {Cincio} \emph
  {et~al.}}]{cerezo2021variational}%
  \BibitemOpen
  \bibfield  {author} {\bibinfo {author} {\bibfnamefont {M.}~\bibnamefont
  {Cerezo}}, \bibinfo {author} {\bibfnamefont {A.}~\bibnamefont {Arrasmith}},
  \bibinfo {author} {\bibfnamefont {R.}~\bibnamefont {Babbush}}, \bibinfo
  {author} {\bibfnamefont {S.~C.}\ \bibnamefont {Benjamin}}, \bibinfo {author}
  {\bibfnamefont {S.}~\bibnamefont {Endo}}, \bibinfo {author} {\bibfnamefont
  {K.}~\bibnamefont {Fujii}}, \bibinfo {author} {\bibfnamefont {J.~R.}\
  \bibnamefont {McClean}}, \bibinfo {author} {\bibfnamefont {K.}~\bibnamefont
  {Mitarai}}, \bibinfo {author} {\bibfnamefont {X.}~\bibnamefont {Yuan}},
  \bibinfo {author} {\bibfnamefont {L.}~\bibnamefont {Cincio}}, \emph
  {et~al.},\ }\bibfield  {title} {\bibinfo {title} {Variational quantum
  algorithms},\ }\href@noop {} {\bibfield  {journal} {\bibinfo  {journal}
  {Nature Reviews Physics}\ ,\ \bibinfo {pages} {1}} (\bibinfo {year}
  {2021})}\BibitemShut {NoStop}%
\bibitem [{\citenamefont {Fedorov}\ \emph {et~al.}(2022)\citenamefont
  {Fedorov}, \citenamefont {Peng}, \citenamefont {Govind},\ and\ \citenamefont
  {Alexeev}}]{fedorov2022vqe}%
  \BibitemOpen
  \bibfield  {author} {\bibinfo {author} {\bibfnamefont {D.~A.}\ \bibnamefont
  {Fedorov}}, \bibinfo {author} {\bibfnamefont {B.}~\bibnamefont {Peng}},
  \bibinfo {author} {\bibfnamefont {N.}~\bibnamefont {Govind}},\ and\ \bibinfo
  {author} {\bibfnamefont {Y.}~\bibnamefont {Alexeev}},\ }\bibfield  {title}
  {\bibinfo {title} {Vqe method: A short survey and recent developments},\
  }\href@noop {} {\bibfield  {journal} {\bibinfo  {journal} {Materials Theory}\
  }\textbf {\bibinfo {volume} {6}},\ \bibinfo {pages} {1} (\bibinfo {year}
  {2022})}\BibitemShut {NoStop}%
\bibitem [{\citenamefont {Cao}\ \emph {et~al.}(2019)\citenamefont {Cao},
  \citenamefont {Romero}, \citenamefont {Olson}, \citenamefont {Degroote},
  \citenamefont {Johnson}, \citenamefont {Kieferov{\'a}}, \citenamefont
  {Kivlichan}, \citenamefont {Menke}, \citenamefont {Peropadre}, \citenamefont
  {Sawaya} \emph {et~al.}}]{cao2019quantum}%
  \BibitemOpen
  \bibfield  {author} {\bibinfo {author} {\bibfnamefont {Y.}~\bibnamefont
  {Cao}}, \bibinfo {author} {\bibfnamefont {J.}~\bibnamefont {Romero}},
  \bibinfo {author} {\bibfnamefont {J.~P.}\ \bibnamefont {Olson}}, \bibinfo
  {author} {\bibfnamefont {M.}~\bibnamefont {Degroote}}, \bibinfo {author}
  {\bibfnamefont {P.~D.}\ \bibnamefont {Johnson}}, \bibinfo {author}
  {\bibfnamefont {M.}~\bibnamefont {Kieferov{\'a}}}, \bibinfo {author}
  {\bibfnamefont {I.~D.}\ \bibnamefont {Kivlichan}}, \bibinfo {author}
  {\bibfnamefont {T.}~\bibnamefont {Menke}}, \bibinfo {author} {\bibfnamefont
  {B.}~\bibnamefont {Peropadre}}, \bibinfo {author} {\bibfnamefont {N.~P.}\
  \bibnamefont {Sawaya}}, \emph {et~al.},\ }\bibfield  {title} {\bibinfo
  {title} {Quantum chemistry in the age of quantum computing},\ }\href@noop {}
  {\bibfield  {journal} {\bibinfo  {journal} {Chemical reviews}\ }\textbf
  {\bibinfo {volume} {119}},\ \bibinfo {pages} {10856} (\bibinfo {year}
  {2019})}\BibitemShut {NoStop}%
\bibitem [{\citenamefont {McArdle}\ \emph {et~al.}(2020)\citenamefont
  {McArdle}, \citenamefont {Endo}, \citenamefont {Aspuru-Guzik}, \citenamefont
  {Benjamin},\ and\ \citenamefont {Yuan}}]{mcardle2020quantum}%
  \BibitemOpen
  \bibfield  {author} {\bibinfo {author} {\bibfnamefont {S.}~\bibnamefont
  {McArdle}}, \bibinfo {author} {\bibfnamefont {S.}~\bibnamefont {Endo}},
  \bibinfo {author} {\bibfnamefont {A.}~\bibnamefont {Aspuru-Guzik}}, \bibinfo
  {author} {\bibfnamefont {S.~C.}\ \bibnamefont {Benjamin}},\ and\ \bibinfo
  {author} {\bibfnamefont {X.}~\bibnamefont {Yuan}},\ }\bibfield  {title}
  {\bibinfo {title} {Quantum computational chemistry},\ }\href@noop {}
  {\bibfield  {journal} {\bibinfo  {journal} {Reviews of Modern Physics}\
  }\textbf {\bibinfo {volume} {92}},\ \bibinfo {pages} {015003} (\bibinfo
  {year} {2020})}\BibitemShut {NoStop}%
\bibitem [{\citenamefont {Gunlycke}(2021)}]{gunlycke}%
  \BibitemOpen
  \bibfield  {author} {\bibinfo {author} {\bibfnamefont {L.~D.}\ \bibnamefont
  {Gunlycke}},\ }\href@noop {} {\bibinfo {title} {Compact, symmetry-adapted
  mapping between fermionic systems and quantum computers}} (\bibinfo {year}
  {U.S. Patent Application 17,184,516, filed Feb. 2021})\BibitemShut {NoStop}%
\bibitem [{\citenamefont {Fischer}\ and\ \citenamefont
  {Gunlycke}(2019)}]{fischer2019symmetry}%
  \BibitemOpen
  \bibfield  {author} {\bibinfo {author} {\bibfnamefont {S.~A.}\ \bibnamefont
  {Fischer}}\ and\ \bibinfo {author} {\bibfnamefont {D.}~\bibnamefont
  {Gunlycke}},\ }\bibfield  {title} {\bibinfo {title} {Symmetry configuration
  mapping for representing quantum systems on quantum computers},\ }\href@noop
  {} {\bibfield  {journal} {\bibinfo  {journal} {arXiv preprint
  arXiv:1907.01493}\ } (\bibinfo {year} {2019})}\BibitemShut {NoStop}%
\bibitem [{\citenamefont {Steudtner}\ and\ \citenamefont
  {Wehner}(2018)}]{steudtner2018fermion}%
  \BibitemOpen
  \bibfield  {author} {\bibinfo {author} {\bibfnamefont {M.}~\bibnamefont
  {Steudtner}}\ and\ \bibinfo {author} {\bibfnamefont {S.}~\bibnamefont
  {Wehner}},\ }\bibfield  {title} {\bibinfo {title} {Fermion-to-qubit mappings
  with varying resource requirements for quantum simulation},\ }\href@noop {}
  {\bibfield  {journal} {\bibinfo  {journal} {New Journal of Physics}\ }\textbf
  {\bibinfo {volume} {20}},\ \bibinfo {pages} {063010} (\bibinfo {year}
  {2018})}\BibitemShut {NoStop}%
\bibitem [{\citenamefont {Steudtner}(2019)}]{steudtner2019methods}%
  \BibitemOpen
  \bibfield  {author} {\bibinfo {author} {\bibfnamefont {M.}~\bibnamefont
  {Steudtner}},\ }\emph {\bibinfo {title} {Methods to simulate fermions on
  quantum computers with hardware limitations}},\ \href@noop {} {Ph.D.
  thesis},\ \bibinfo  {school} {Leiden} (\bibinfo {year} {2019})\BibitemShut
  {NoStop}%
\bibitem [{\citenamefont {Chamaki}\ \emph {et~al.}(2022)\citenamefont
  {Chamaki}, \citenamefont {Metcalf},\ and\ \citenamefont
  {de~Jong}}]{chamaki2022compact}%
  \BibitemOpen
  \bibfield  {author} {\bibinfo {author} {\bibfnamefont {D.}~\bibnamefont
  {Chamaki}}, \bibinfo {author} {\bibfnamefont {M.}~\bibnamefont {Metcalf}},\
  and\ \bibinfo {author} {\bibfnamefont {W.~A.}\ \bibnamefont {de~Jong}},\
  }\bibfield  {title} {\bibinfo {title} {Compact molecular simulation on
  quantum computers via combinatorial mapping and variational state
  preparation},\ }\href@noop {} {\bibfield  {journal} {\bibinfo  {journal}
  {arXiv preprint arXiv:2205.11742}\ } (\bibinfo {year} {2022})}\BibitemShut
  {NoStop}%
\bibitem [{\citenamefont {Shee}\ \emph {et~al.}(2022)\citenamefont {Shee},
  \citenamefont {Tsai}, \citenamefont {Hong}, \citenamefont {Cheng},\ and\
  \citenamefont {Goan}}]{shee2022qubit}%
  \BibitemOpen
  \bibfield  {author} {\bibinfo {author} {\bibfnamefont {Y.}~\bibnamefont
  {Shee}}, \bibinfo {author} {\bibfnamefont {P.-K.}\ \bibnamefont {Tsai}},
  \bibinfo {author} {\bibfnamefont {C.-L.}\ \bibnamefont {Hong}}, \bibinfo
  {author} {\bibfnamefont {H.-C.}\ \bibnamefont {Cheng}},\ and\ \bibinfo
  {author} {\bibfnamefont {H.-S.}\ \bibnamefont {Goan}},\ }\bibfield  {title}
  {\bibinfo {title} {Qubit-efficient encoding scheme for quantum simulations of
  electronic structure},\ }\href@noop {} {\bibfield  {journal} {\bibinfo
  {journal} {Physical Review Research}\ }\textbf {\bibinfo {volume} {4}},\
  \bibinfo {pages} {023154} (\bibinfo {year} {2022})}\BibitemShut {NoStop}%
\bibitem [{\citenamefont {Romero}\ \emph {et~al.}(2018)\citenamefont {Romero},
  \citenamefont {Babbush}, \citenamefont {McClean}, \citenamefont {Hempel},
  \citenamefont {Love},\ and\ \citenamefont
  {Aspuru-Guzik}}]{romero2018strategies}%
  \BibitemOpen
  \bibfield  {author} {\bibinfo {author} {\bibfnamefont {J.}~\bibnamefont
  {Romero}}, \bibinfo {author} {\bibfnamefont {R.}~\bibnamefont {Babbush}},
  \bibinfo {author} {\bibfnamefont {J.~R.}\ \bibnamefont {McClean}}, \bibinfo
  {author} {\bibfnamefont {C.}~\bibnamefont {Hempel}}, \bibinfo {author}
  {\bibfnamefont {P.~J.}\ \bibnamefont {Love}},\ and\ \bibinfo {author}
  {\bibfnamefont {A.}~\bibnamefont {Aspuru-Guzik}},\ }\bibfield  {title}
  {\bibinfo {title} {Strategies for quantum computing molecular energies using
  the unitary coupled cluster ansatz},\ }\href@noop {} {\bibfield  {journal}
  {\bibinfo  {journal} {Quantum Science and Technology}\ }\textbf {\bibinfo
  {volume} {4}},\ \bibinfo {pages} {014008} (\bibinfo {year}
  {2018})}\BibitemShut {NoStop}%
\bibitem [{\citenamefont {Grimsley}\ \emph {et~al.}(2019)\citenamefont
  {Grimsley}, \citenamefont {Economou}, \citenamefont {Barnes},\ and\
  \citenamefont {Mayhall}}]{grimsley2019adaptive}%
  \BibitemOpen
  \bibfield  {author} {\bibinfo {author} {\bibfnamefont {H.~R.}\ \bibnamefont
  {Grimsley}}, \bibinfo {author} {\bibfnamefont {S.~E.}\ \bibnamefont
  {Economou}}, \bibinfo {author} {\bibfnamefont {E.}~\bibnamefont {Barnes}},\
  and\ \bibinfo {author} {\bibfnamefont {N.~J.}\ \bibnamefont {Mayhall}},\
  }\bibfield  {title} {\bibinfo {title} {An adaptive variational algorithm for
  exact molecular simulations on a quantum computer},\ }\href@noop {}
  {\bibfield  {journal} {\bibinfo  {journal} {Nature communications}\ }\textbf
  {\bibinfo {volume} {10}},\ \bibinfo {pages} {1} (\bibinfo {year}
  {2019})}\BibitemShut {NoStop}%
\bibitem [{\citenamefont {Anselmetti}\ \emph {et~al.}(2021)\citenamefont
  {Anselmetti}, \citenamefont {Wierichs}, \citenamefont {Gogolin},\ and\
  \citenamefont {Parrish}}]{anselmetti2021local}%
  \BibitemOpen
  \bibfield  {author} {\bibinfo {author} {\bibfnamefont {G.-L.~R.}\
  \bibnamefont {Anselmetti}}, \bibinfo {author} {\bibfnamefont
  {D.}~\bibnamefont {Wierichs}}, \bibinfo {author} {\bibfnamefont
  {C.}~\bibnamefont {Gogolin}},\ and\ \bibinfo {author} {\bibfnamefont {R.~M.}\
  \bibnamefont {Parrish}},\ }\bibfield  {title} {\bibinfo {title} {Local,
  expressive, quantum-number-preserving vqe ansatze for fermionic systems},\
  }\href@noop {} {\bibfield  {journal} {\bibinfo  {journal} {arXiv preprint
  arXiv:2104.05695}\ } (\bibinfo {year} {2021})}\BibitemShut {NoStop}%
\bibitem [{\citenamefont {Gard}\ \emph {et~al.}(2020)\citenamefont {Gard},
  \citenamefont {Zhu}, \citenamefont {Barron}, \citenamefont {Mayhall},
  \citenamefont {Economou},\ and\ \citenamefont {Barnes}}]{gard2020efficient}%
  \BibitemOpen
  \bibfield  {author} {\bibinfo {author} {\bibfnamefont {B.~T.}\ \bibnamefont
  {Gard}}, \bibinfo {author} {\bibfnamefont {L.}~\bibnamefont {Zhu}}, \bibinfo
  {author} {\bibfnamefont {G.~S.}\ \bibnamefont {Barron}}, \bibinfo {author}
  {\bibfnamefont {N.~J.}\ \bibnamefont {Mayhall}}, \bibinfo {author}
  {\bibfnamefont {S.~E.}\ \bibnamefont {Economou}},\ and\ \bibinfo {author}
  {\bibfnamefont {E.}~\bibnamefont {Barnes}},\ }\bibfield  {title} {\bibinfo
  {title} {Efficient symmetry-preserving state preparation circuits for the
  variational quantum eigensolver algorithm},\ }\href@noop {} {\bibfield
  {journal} {\bibinfo  {journal} {npj Quantum Information}\ }\textbf {\bibinfo
  {volume} {6}},\ \bibinfo {pages} {1} (\bibinfo {year} {2020})}\BibitemShut
  {NoStop}%
\bibitem [{\citenamefont {McClean}\ \emph {et~al.}(2016)\citenamefont
  {McClean}, \citenamefont {Romero}, \citenamefont {Babbush},\ and\
  \citenamefont {Aspuru-Guzik}}]{mcclean2016theory}%
  \BibitemOpen
  \bibfield  {author} {\bibinfo {author} {\bibfnamefont {J.~R.}\ \bibnamefont
  {McClean}}, \bibinfo {author} {\bibfnamefont {J.}~\bibnamefont {Romero}},
  \bibinfo {author} {\bibfnamefont {R.}~\bibnamefont {Babbush}},\ and\ \bibinfo
  {author} {\bibfnamefont {A.}~\bibnamefont {Aspuru-Guzik}},\ }\bibfield
  {title} {\bibinfo {title} {The theory of variational hybrid quantum-classical
  algorithms},\ }\href@noop {} {\bibfield  {journal} {\bibinfo  {journal} {New
  Journal of Physics}\ }\textbf {\bibinfo {volume} {18}},\ \bibinfo {pages}
  {023023} (\bibinfo {year} {2016})}\BibitemShut {NoStop}%
\bibitem [{\citenamefont {Ryabinkin}\ \emph {et~al.}(2018)\citenamefont
  {Ryabinkin}, \citenamefont {Genin},\ and\ \citenamefont
  {Izmaylov}}]{ryabinkin2018constrained}%
  \BibitemOpen
  \bibfield  {author} {\bibinfo {author} {\bibfnamefont {I.~G.}\ \bibnamefont
  {Ryabinkin}}, \bibinfo {author} {\bibfnamefont {S.~N.}\ \bibnamefont
  {Genin}},\ and\ \bibinfo {author} {\bibfnamefont {A.~F.}\ \bibnamefont
  {Izmaylov}},\ }\bibfield  {title} {\bibinfo {title} {Constrained variational
  quantum eigensolver: Quantum computer search engine in the fock space},\
  }\href@noop {} {\bibfield  {journal} {\bibinfo  {journal} {Journal of
  chemical theory and computation}\ }\textbf {\bibinfo {volume} {15}},\
  \bibinfo {pages} {249} (\bibinfo {year} {2018})}\BibitemShut {NoStop}%
\bibitem [{\citenamefont {Kuroiwa}\ and\ \citenamefont
  {Nakagawa}(2021)}]{kuroiwa2021penalty}%
  \BibitemOpen
  \bibfield  {author} {\bibinfo {author} {\bibfnamefont {K.}~\bibnamefont
  {Kuroiwa}}\ and\ \bibinfo {author} {\bibfnamefont {Y.~O.}\ \bibnamefont
  {Nakagawa}},\ }\bibfield  {title} {\bibinfo {title} {Penalty methods for a
  variational quantum eigensolver},\ }\href@noop {} {\bibfield  {journal}
  {\bibinfo  {journal} {Physical Review Research}\ }\textbf {\bibinfo {volume}
  {3}},\ \bibinfo {pages} {013197} (\bibinfo {year} {2021})}\BibitemShut
  {NoStop}%
\bibitem [{\citenamefont {Nielsen}\ \emph {et~al.}(2005)\citenamefont {Nielsen}
  \emph {et~al.}}]{nielsen2005fermionic}%
  \BibitemOpen
  \bibfield  {author} {\bibinfo {author} {\bibfnamefont {M.~A.}\ \bibnamefont
  {Nielsen}} \emph {et~al.},\ }\bibfield  {title} {\bibinfo {title} {The
  fermionic canonical commutation relations and the jordan-wigner transform},\
  }\href@noop {} {\bibfield  {journal} {\bibinfo  {journal} {School of Physical
  Sciences The University of Queensland}\ }\textbf {\bibinfo {volume} {59}}
  (\bibinfo {year} {2005})}\BibitemShut {NoStop}%
\bibitem [{\citenamefont {Jordan}\ and\ \citenamefont
  {Wigner}(1993)}]{jordan1993paulische}%
  \BibitemOpen
  \bibfield  {author} {\bibinfo {author} {\bibfnamefont {P.}~\bibnamefont
  {Jordan}}\ and\ \bibinfo {author} {\bibfnamefont {E.~P.}\ \bibnamefont
  {Wigner}},\ }\bibfield  {title} {\bibinfo {title} {{\"u}ber das paulische
  {\"a}quivalenzverbot},\ }in\ \href@noop {} {\emph {\bibinfo {booktitle} {The
  Collected Works of Eugene Paul Wigner}}}\ (\bibinfo  {publisher} {Springer},\
  \bibinfo {year} {1993})\ pp.\ \bibinfo {pages} {109--129}\BibitemShut
  {NoStop}%
\bibitem [{\citenamefont {Seeley}\ \emph {et~al.}(2012)\citenamefont {Seeley},
  \citenamefont {Richard},\ and\ \citenamefont {Love}}]{seeley2012bravyi}%
  \BibitemOpen
  \bibfield  {author} {\bibinfo {author} {\bibfnamefont {J.~T.}\ \bibnamefont
  {Seeley}}, \bibinfo {author} {\bibfnamefont {M.~J.}\ \bibnamefont
  {Richard}},\ and\ \bibinfo {author} {\bibfnamefont {P.~J.}\ \bibnamefont
  {Love}},\ }\bibfield  {title} {\bibinfo {title} {The bravyi-kitaev
  transformation for quantum computation of electronic structure},\ }\href@noop
  {} {\bibfield  {journal} {\bibinfo  {journal} {The Journal of chemical
  physics}\ }\textbf {\bibinfo {volume} {137}},\ \bibinfo {pages} {224109}
  (\bibinfo {year} {2012})}\BibitemShut {NoStop}%
\bibitem [{\citenamefont {Setia}\ and\ \citenamefont
  {Whitfield}(2018)}]{setia2018bravyi}%
  \BibitemOpen
  \bibfield  {author} {\bibinfo {author} {\bibfnamefont {K.}~\bibnamefont
  {Setia}}\ and\ \bibinfo {author} {\bibfnamefont {J.~D.}\ \bibnamefont
  {Whitfield}},\ }\bibfield  {title} {\bibinfo {title} {Bravyi-kitaev superfast
  simulation of electronic structure on a quantum computer},\ }\href@noop {}
  {\bibfield  {journal} {\bibinfo  {journal} {The Journal of chemical physics}\
  }\textbf {\bibinfo {volume} {148}},\ \bibinfo {pages} {164104} (\bibinfo
  {year} {2018})}\BibitemShut {NoStop}%
\bibitem [{\citenamefont {et~al.}(2020)}]{mezz}%
  \BibitemOpen
  \bibfield  {author} {\bibinfo {author} {\bibfnamefont {M.}~\bibnamefont
  {et~al.}},\ }\href@noop {} {\bibinfo {title} {Hardware-efficient variational
  quantum eigenvalue solver for quantum computing machines}} (\bibinfo {year}
  {U.S. Patent 10,839,306, Nov. 2020})\BibitemShut {NoStop}%
\bibitem [{\citenamefont {Kandala}\ \emph {et~al.}(2017)\citenamefont
  {Kandala}, \citenamefont {Mezzacapo}, \citenamefont {Temme}, \citenamefont
  {Takita}, \citenamefont {Brink}, \citenamefont {Chow},\ and\ \citenamefont
  {Gambetta}}]{kandala2017hardware}%
  \BibitemOpen
  \bibfield  {author} {\bibinfo {author} {\bibfnamefont {A.}~\bibnamefont
  {Kandala}}, \bibinfo {author} {\bibfnamefont {A.}~\bibnamefont {Mezzacapo}},
  \bibinfo {author} {\bibfnamefont {K.}~\bibnamefont {Temme}}, \bibinfo
  {author} {\bibfnamefont {M.}~\bibnamefont {Takita}}, \bibinfo {author}
  {\bibfnamefont {M.}~\bibnamefont {Brink}}, \bibinfo {author} {\bibfnamefont
  {J.~M.}\ \bibnamefont {Chow}},\ and\ \bibinfo {author} {\bibfnamefont
  {J.~M.}\ \bibnamefont {Gambetta}},\ }\bibfield  {title} {\bibinfo {title}
  {Hardware-efficient variational quantum eigensolver for small molecules and
  quantum magnets},\ }\href@noop {} {\bibfield  {journal} {\bibinfo  {journal}
  {Nature}\ }\textbf {\bibinfo {volume} {549}},\ \bibinfo {pages} {242}
  (\bibinfo {year} {2017})}\BibitemShut {NoStop}%
\bibitem [{\citenamefont {O'Gorman}\ \emph {et~al.}(2019)\citenamefont
  {O'Gorman}, \citenamefont {Huggins}, \citenamefont {Rieffel},\ and\
  \citenamefont {Whaley}}]{o2019generalized}%
  \BibitemOpen
  \bibfield  {author} {\bibinfo {author} {\bibfnamefont {B.}~\bibnamefont
  {O'Gorman}}, \bibinfo {author} {\bibfnamefont {W.~J.}\ \bibnamefont
  {Huggins}}, \bibinfo {author} {\bibfnamefont {E.~G.}\ \bibnamefont
  {Rieffel}},\ and\ \bibinfo {author} {\bibfnamefont {K.~B.}\ \bibnamefont
  {Whaley}},\ }\bibfield  {title} {\bibinfo {title} {Generalized swap networks
  for near-term quantum computing},\ }\href@noop {} {\bibfield  {journal}
  {\bibinfo  {journal} {arXiv preprint arXiv:1905.05118}\ } (\bibinfo {year}
  {2019})}\BibitemShut {NoStop}%
\bibitem [{\citenamefont {Koridon}\ \emph {et~al.}(2021)\citenamefont
  {Koridon}, \citenamefont {Yalouz}, \citenamefont {Senjean}, \citenamefont
  {Buda}, \citenamefont {O'Brien},\ and\ \citenamefont
  {Visscher}}]{koridon2021orbital}%
  \BibitemOpen
  \bibfield  {author} {\bibinfo {author} {\bibfnamefont {E.}~\bibnamefont
  {Koridon}}, \bibinfo {author} {\bibfnamefont {S.}~\bibnamefont {Yalouz}},
  \bibinfo {author} {\bibfnamefont {B.}~\bibnamefont {Senjean}}, \bibinfo
  {author} {\bibfnamefont {F.}~\bibnamefont {Buda}}, \bibinfo {author}
  {\bibfnamefont {T.~E.}\ \bibnamefont {O'Brien}},\ and\ \bibinfo {author}
  {\bibfnamefont {L.}~\bibnamefont {Visscher}},\ }\bibfield  {title} {\bibinfo
  {title} {Orbital transformations to reduce the 1-norm of the electronic
  structure hamiltonian for quantum computing applications},\ }\href@noop {}
  {\bibfield  {journal} {\bibinfo  {journal} {Physical Review Research}\
  }\textbf {\bibinfo {volume} {3}},\ \bibinfo {pages} {033127} (\bibinfo {year}
  {2021})}\BibitemShut {NoStop}%
\bibitem [{\citenamefont {Tsuchimochi}\ \emph {et~al.}(2022)\citenamefont
  {Tsuchimochi}, \citenamefont {Taii}, \citenamefont {Nishimaki},\ and\
  \citenamefont {Ten-no}}]{tsuchimochi2022adaptive}%
  \BibitemOpen
  \bibfield  {author} {\bibinfo {author} {\bibfnamefont {T.}~\bibnamefont
  {Tsuchimochi}}, \bibinfo {author} {\bibfnamefont {M.}~\bibnamefont {Taii}},
  \bibinfo {author} {\bibfnamefont {T.}~\bibnamefont {Nishimaki}},\ and\
  \bibinfo {author} {\bibfnamefont {S.~L.}\ \bibnamefont {Ten-no}},\ }\bibfield
   {title} {\bibinfo {title} {Adaptive construction of shallower quantum
  circuits with quantum spin projection for fermionic systems},\ }\href@noop {}
  {\bibfield  {journal} {\bibinfo  {journal} {arXiv preprint arXiv:2205.07097}\
  } (\bibinfo {year} {2022})}\BibitemShut {NoStop}%
\bibitem [{\citenamefont {Li}\ \emph {et~al.}(2022)\citenamefont {Li},
  \citenamefont {Huang}, \citenamefont {Cao}, \citenamefont {Huang},
  \citenamefont {Shuai}, \citenamefont {Sun}, \citenamefont {Sun},
  \citenamefont {Yuan},\ and\ \citenamefont {Lv}}]{li2022toward}%
  \BibitemOpen
  \bibfield  {author} {\bibinfo {author} {\bibfnamefont {W.}~\bibnamefont
  {Li}}, \bibinfo {author} {\bibfnamefont {Z.}~\bibnamefont {Huang}}, \bibinfo
  {author} {\bibfnamefont {C.}~\bibnamefont {Cao}}, \bibinfo {author}
  {\bibfnamefont {Y.}~\bibnamefont {Huang}}, \bibinfo {author} {\bibfnamefont
  {Z.}~\bibnamefont {Shuai}}, \bibinfo {author} {\bibfnamefont
  {X.}~\bibnamefont {Sun}}, \bibinfo {author} {\bibfnamefont {J.}~\bibnamefont
  {Sun}}, \bibinfo {author} {\bibfnamefont {X.}~\bibnamefont {Yuan}},\ and\
  \bibinfo {author} {\bibfnamefont {D.}~\bibnamefont {Lv}},\ }\bibfield
  {title} {\bibinfo {title} {Toward practical quantum embedding simulation of
  realistic chemical systems on near-term quantum computers},\ }\href@noop {}
  {\bibfield  {journal} {\bibinfo  {journal} {Chemical science}\ }\textbf
  {\bibinfo {volume} {13}},\ \bibinfo {pages} {8953} (\bibinfo {year}
  {2022})}\BibitemShut {NoStop}%
\bibitem [{\citenamefont {Elfving}\ \emph {et~al.}(2020)\citenamefont
  {Elfving}, \citenamefont {Broer}, \citenamefont {Webber}, \citenamefont
  {Gavartin}, \citenamefont {Halls}, \citenamefont {Lorton},\ and\
  \citenamefont {Bochevarov}}]{elfving2020will}%
  \BibitemOpen
  \bibfield  {author} {\bibinfo {author} {\bibfnamefont {V.~E.}\ \bibnamefont
  {Elfving}}, \bibinfo {author} {\bibfnamefont {B.~W.}\ \bibnamefont {Broer}},
  \bibinfo {author} {\bibfnamefont {M.}~\bibnamefont {Webber}}, \bibinfo
  {author} {\bibfnamefont {J.}~\bibnamefont {Gavartin}}, \bibinfo {author}
  {\bibfnamefont {M.~D.}\ \bibnamefont {Halls}}, \bibinfo {author}
  {\bibfnamefont {K.~P.}\ \bibnamefont {Lorton}},\ and\ \bibinfo {author}
  {\bibfnamefont {A.}~\bibnamefont {Bochevarov}},\ }\bibfield  {title}
  {\bibinfo {title} {How will quantum computers provide an industrially
  relevant computational advantage in quantum chemistry?},\ }\href@noop {}
  {\bibfield  {journal} {\bibinfo  {journal} {arXiv preprint arXiv:2009.12472}\
  } (\bibinfo {year} {2020})}\BibitemShut {NoStop}%
\bibitem [{\citenamefont {Fujii}\ \emph {et~al.}(2022)\citenamefont {Fujii},
  \citenamefont {Mizuta}, \citenamefont {Ueda}, \citenamefont {Mitarai},
  \citenamefont {Mizukami},\ and\ \citenamefont {Nakagawa}}]{fujii2022deep}%
  \BibitemOpen
  \bibfield  {author} {\bibinfo {author} {\bibfnamefont {K.}~\bibnamefont
  {Fujii}}, \bibinfo {author} {\bibfnamefont {K.}~\bibnamefont {Mizuta}},
  \bibinfo {author} {\bibfnamefont {H.}~\bibnamefont {Ueda}}, \bibinfo {author}
  {\bibfnamefont {K.}~\bibnamefont {Mitarai}}, \bibinfo {author} {\bibfnamefont
  {W.}~\bibnamefont {Mizukami}},\ and\ \bibinfo {author} {\bibfnamefont
  {Y.~O.}\ \bibnamefont {Nakagawa}},\ }\bibfield  {title} {\bibinfo {title}
  {Deep variational quantum eigensolver: a divide-and-conquer method for
  solving a larger problem with smaller size quantum computers},\ }\href@noop
  {} {\bibfield  {journal} {\bibinfo  {journal} {PRX Quantum}\ }\textbf
  {\bibinfo {volume} {3}},\ \bibinfo {pages} {010346} (\bibinfo {year}
  {2022})}\BibitemShut {NoStop}%
\bibitem [{\citenamefont {Kumar}\ \emph {et~al.}(2022)\citenamefont {Kumar},
  \citenamefont {Asthana}, \citenamefont {Masteran}, \citenamefont {Valeev},
  \citenamefont {Zhang}, \citenamefont {Cincio}, \citenamefont {Tretiak},\ and\
  \citenamefont {Dub}}]{kumar2022quantum}%
  \BibitemOpen
  \bibfield  {author} {\bibinfo {author} {\bibfnamefont {A.}~\bibnamefont
  {Kumar}}, \bibinfo {author} {\bibfnamefont {A.}~\bibnamefont {Asthana}},
  \bibinfo {author} {\bibfnamefont {C.}~\bibnamefont {Masteran}}, \bibinfo
  {author} {\bibfnamefont {E.~F.}\ \bibnamefont {Valeev}}, \bibinfo {author}
  {\bibfnamefont {Y.}~\bibnamefont {Zhang}}, \bibinfo {author} {\bibfnamefont
  {L.}~\bibnamefont {Cincio}}, \bibinfo {author} {\bibfnamefont
  {S.}~\bibnamefont {Tretiak}},\ and\ \bibinfo {author} {\bibfnamefont {P.~A.}\
  \bibnamefont {Dub}},\ }\bibfield  {title} {\bibinfo {title} {Quantum
  simulation of molecular electronic states with a transcorrelated hamiltonian:
  Higher accuracy with fewer qubits},\ }\href@noop {} {\bibfield  {journal}
  {\bibinfo  {journal} {Journal of Chemical Theory and Computation}\ }\textbf
  {\bibinfo {volume} {18}},\ \bibinfo {pages} {5312} (\bibinfo {year}
  {2022})}\BibitemShut {NoStop}%
\bibitem [{\citenamefont {Bauman}\ and\ \citenamefont
  {Kowalski}(2022)}]{bauman2022coupled}%
  \BibitemOpen
  \bibfield  {author} {\bibinfo {author} {\bibfnamefont {N.~P.}\ \bibnamefont
  {Bauman}}\ and\ \bibinfo {author} {\bibfnamefont {K.}~\bibnamefont
  {Kowalski}},\ }\bibfield  {title} {\bibinfo {title} {Coupled cluster
  downfolding methods: The effect of double commutator terms on the accuracy of
  ground-state energies},\ }\href@noop {} {\bibfield  {journal} {\bibinfo
  {journal} {The Journal of Chemical Physics}\ }\textbf {\bibinfo {volume}
  {156}},\ \bibinfo {pages} {094106} (\bibinfo {year} {2022})}\BibitemShut
  {NoStop}%
\bibitem [{\citenamefont {Dhawan}\ \emph {et~al.}(2020)\citenamefont {Dhawan},
  \citenamefont {Metcalf},\ and\ \citenamefont {Zgid}}]{dhawan2020dynamical}%
  \BibitemOpen
  \bibfield  {author} {\bibinfo {author} {\bibfnamefont {D.}~\bibnamefont
  {Dhawan}}, \bibinfo {author} {\bibfnamefont {M.}~\bibnamefont {Metcalf}},\
  and\ \bibinfo {author} {\bibfnamefont {D.}~\bibnamefont {Zgid}},\ }\bibfield
  {title} {\bibinfo {title} {Dynamical self-energy mapping (dsem) for quantum
  computing},\ }\href@noop {} {\bibfield  {journal} {\bibinfo  {journal} {arXiv
  preprint arXiv:2010.05441}\ } (\bibinfo {year} {2020})}\BibitemShut {NoStop}%
\bibitem [{\citenamefont {et~al.}(2019)}]{taper}%
  \BibitemOpen
  \bibfield  {author} {\bibinfo {author} {\bibfnamefont {K.~S.}\ \bibnamefont
  {et~al.}},\ }\href@noop {} {\bibinfo {title} {Precision-preserving qubit
  reduction based on spatial symmetries in fermionic systems}} (\bibinfo {year}
  {U.S. Patent Application 16,660,059, filed Oct. 2019})\BibitemShut {NoStop}%
\bibitem [{\citenamefont {McClean}\ \emph {et~al.}(2017)\citenamefont
  {McClean}, \citenamefont {Kimchi-Schwartz}, \citenamefont {Carter},\ and\
  \citenamefont {De~Jong}}]{mcclean2017hybrid}%
  \BibitemOpen
  \bibfield  {author} {\bibinfo {author} {\bibfnamefont {J.~R.}\ \bibnamefont
  {McClean}}, \bibinfo {author} {\bibfnamefont {M.~E.}\ \bibnamefont
  {Kimchi-Schwartz}}, \bibinfo {author} {\bibfnamefont {J.}~\bibnamefont
  {Carter}},\ and\ \bibinfo {author} {\bibfnamefont {W.~A.}\ \bibnamefont
  {De~Jong}},\ }\bibfield  {title} {\bibinfo {title} {Hybrid quantum-classical
  hierarchy for mitigation of decoherence and determination of excited
  states},\ }\href@noop {} {\bibfield  {journal} {\bibinfo  {journal} {Physical
  Review A}\ }\textbf {\bibinfo {volume} {95}},\ \bibinfo {pages} {042308}
  (\bibinfo {year} {2017})}\BibitemShut {NoStop}%
\bibitem [{\citenamefont {Urbanek}\ \emph {et~al.}(2020)\citenamefont
  {Urbanek}, \citenamefont {Camps}, \citenamefont {Van~Beeumen},\ and\
  \citenamefont {de~Jong}}]{urbanek2020chemistry}%
  \BibitemOpen
  \bibfield  {author} {\bibinfo {author} {\bibfnamefont {M.}~\bibnamefont
  {Urbanek}}, \bibinfo {author} {\bibfnamefont {D.}~\bibnamefont {Camps}},
  \bibinfo {author} {\bibfnamefont {R.}~\bibnamefont {Van~Beeumen}},\ and\
  \bibinfo {author} {\bibfnamefont {W.~A.}\ \bibnamefont {de~Jong}},\
  }\bibfield  {title} {\bibinfo {title} {Chemistry on quantum computers with
  virtual quantum subspace expansion},\ }\href@noop {} {\bibfield  {journal}
  {\bibinfo  {journal} {Journal of chemical theory and computation}\ }\textbf
  {\bibinfo {volume} {16}},\ \bibinfo {pages} {5425} (\bibinfo {year}
  {2020})}\BibitemShut {NoStop}%
\bibitem [{\citenamefont {Verteletskyi}\ \emph {et~al.}(2020)\citenamefont
  {Verteletskyi}, \citenamefont {Yen},\ and\ \citenamefont
  {Izmaylov}}]{verteletskyi2020measurement}%
  \BibitemOpen
  \bibfield  {author} {\bibinfo {author} {\bibfnamefont {V.}~\bibnamefont
  {Verteletskyi}}, \bibinfo {author} {\bibfnamefont {T.-C.}\ \bibnamefont
  {Yen}},\ and\ \bibinfo {author} {\bibfnamefont {A.~F.}\ \bibnamefont
  {Izmaylov}},\ }\bibfield  {title} {\bibinfo {title} {Measurement optimization
  in the variational quantum eigensolver using a minimum clique cover},\
  }\href@noop {} {\bibfield  {journal} {\bibinfo  {journal} {The Journal of
  chemical physics}\ }\textbf {\bibinfo {volume} {152}},\ \bibinfo {pages}
  {124114} (\bibinfo {year} {2020})}\BibitemShut {NoStop}%
\bibitem [{\citenamefont {Yen}\ \emph {et~al.}(2020)\citenamefont {Yen},
  \citenamefont {Verteletskyi},\ and\ \citenamefont
  {Izmaylov}}]{yen2020measuring}%
  \BibitemOpen
  \bibfield  {author} {\bibinfo {author} {\bibfnamefont {T.-C.}\ \bibnamefont
  {Yen}}, \bibinfo {author} {\bibfnamefont {V.}~\bibnamefont {Verteletskyi}},\
  and\ \bibinfo {author} {\bibfnamefont {A.~F.}\ \bibnamefont {Izmaylov}},\
  }\bibfield  {title} {\bibinfo {title} {Measuring all compatible operators in
  one series of single-qubit measurements using unitary transformations},\
  }\href@noop {} {\bibfield  {journal} {\bibinfo  {journal} {Journal of
  chemical theory and computation}\ }\textbf {\bibinfo {volume} {16}},\
  \bibinfo {pages} {2400} (\bibinfo {year} {2020})}\BibitemShut {NoStop}%
\bibitem [{\citenamefont {Izmaylov}\ \emph {et~al.}(2019)\citenamefont
  {Izmaylov}, \citenamefont {Yen}, \citenamefont {Lang},\ and\ \citenamefont
  {Verteletskyi}}]{izmaylov2019unitary}%
  \BibitemOpen
  \bibfield  {author} {\bibinfo {author} {\bibfnamefont {A.~F.}\ \bibnamefont
  {Izmaylov}}, \bibinfo {author} {\bibfnamefont {T.-C.}\ \bibnamefont {Yen}},
  \bibinfo {author} {\bibfnamefont {R.~A.}\ \bibnamefont {Lang}},\ and\
  \bibinfo {author} {\bibfnamefont {V.}~\bibnamefont {Verteletskyi}},\
  }\bibfield  {title} {\bibinfo {title} {Unitary partitioning approach to the
  measurement problem in the variational quantum eigensolver method},\
  }\href@noop {} {\bibfield  {journal} {\bibinfo  {journal} {Journal of
  chemical theory and computation}\ }\textbf {\bibinfo {volume} {16}},\
  \bibinfo {pages} {190} (\bibinfo {year} {2019})}\BibitemShut {NoStop}%
\bibitem [{\citenamefont {Crawford}\ \emph {et~al.}(2021)\citenamefont
  {Crawford}, \citenamefont {van Straaten}, \citenamefont {Wang}, \citenamefont
  {Parks}, \citenamefont {Campbell},\ and\ \citenamefont
  {Brierley}}]{crawford2021efficient}%
  \BibitemOpen
  \bibfield  {author} {\bibinfo {author} {\bibfnamefont {O.}~\bibnamefont
  {Crawford}}, \bibinfo {author} {\bibfnamefont {B.}~\bibnamefont {van
  Straaten}}, \bibinfo {author} {\bibfnamefont {D.}~\bibnamefont {Wang}},
  \bibinfo {author} {\bibfnamefont {T.}~\bibnamefont {Parks}}, \bibinfo
  {author} {\bibfnamefont {E.}~\bibnamefont {Campbell}},\ and\ \bibinfo
  {author} {\bibfnamefont {S.}~\bibnamefont {Brierley}},\ }\bibfield  {title}
  {\bibinfo {title} {Efficient quantum measurement of pauli operators in the
  presence of finite sampling error},\ }\href@noop {} {\bibfield  {journal}
  {\bibinfo  {journal} {Quantum}\ }\textbf {\bibinfo {volume} {5}},\ \bibinfo
  {pages} {385} (\bibinfo {year} {2021})}\BibitemShut {NoStop}%
\bibitem [{\citenamefont {Hamamura}\ and\ \citenamefont
  {Imamichi}(2020)}]{hamamura2020efficient}%
  \BibitemOpen
  \bibfield  {author} {\bibinfo {author} {\bibfnamefont {I.}~\bibnamefont
  {Hamamura}}\ and\ \bibinfo {author} {\bibfnamefont {T.}~\bibnamefont
  {Imamichi}},\ }\bibfield  {title} {\bibinfo {title} {Efficient evaluation of
  quantum observables using entangled measurements},\ }\href@noop {} {\bibfield
   {journal} {\bibinfo  {journal} {npj Quantum Information}\ }\textbf {\bibinfo
  {volume} {6}},\ \bibinfo {pages} {1} (\bibinfo {year} {2020})}\BibitemShut
  {NoStop}%
\bibitem [{\citenamefont {Tang}\ \emph {et~al.}(2021)\citenamefont {Tang},
  \citenamefont {Shkolnikov}, \citenamefont {Barron}, \citenamefont {Grimsley},
  \citenamefont {Mayhall}, \citenamefont {Barnes},\ and\ \citenamefont
  {Economou}}]{tang2021qubit}%
  \BibitemOpen
  \bibfield  {author} {\bibinfo {author} {\bibfnamefont {H.~L.}\ \bibnamefont
  {Tang}}, \bibinfo {author} {\bibfnamefont {V.}~\bibnamefont {Shkolnikov}},
  \bibinfo {author} {\bibfnamefont {G.~S.}\ \bibnamefont {Barron}}, \bibinfo
  {author} {\bibfnamefont {H.~R.}\ \bibnamefont {Grimsley}}, \bibinfo {author}
  {\bibfnamefont {N.~J.}\ \bibnamefont {Mayhall}}, \bibinfo {author}
  {\bibfnamefont {E.}~\bibnamefont {Barnes}},\ and\ \bibinfo {author}
  {\bibfnamefont {S.~E.}\ \bibnamefont {Economou}},\ }\bibfield  {title}
  {\bibinfo {title} {qubit-adapt-vqe: An adaptive algorithm for constructing
  hardware-efficient ans{\"a}tze on a quantum processor},\ }\href@noop {}
  {\bibfield  {journal} {\bibinfo  {journal} {PRX Quantum}\ }\textbf {\bibinfo
  {volume} {2}},\ \bibinfo {pages} {020310} (\bibinfo {year}
  {2021})}\BibitemShut {NoStop}%
\bibitem [{\citenamefont {Helgaker}\ \emph {et~al.}(2014)\citenamefont
  {Helgaker}, \citenamefont {Jorgensen},\ and\ \citenamefont
  {Olsen}}]{helgaker2014molecular}%
  \BibitemOpen
  \bibfield  {author} {\bibinfo {author} {\bibfnamefont {T.}~\bibnamefont
  {Helgaker}}, \bibinfo {author} {\bibfnamefont {P.}~\bibnamefont
  {Jorgensen}},\ and\ \bibinfo {author} {\bibfnamefont {J.}~\bibnamefont
  {Olsen}},\ }\href@noop {} {\emph {\bibinfo {title} {Molecular
  electronic-structure theory}}}\ (\bibinfo  {publisher} {John Wiley \& Sons},\
  \bibinfo {year} {2014})\BibitemShut {NoStop}%
\bibitem [{\citenamefont {Setia}\ \emph {et~al.}(2020)\citenamefont {Setia},
  \citenamefont {Chen}, \citenamefont {Rice}, \citenamefont {Mezzacapo},
  \citenamefont {Pistoia},\ and\ \citenamefont
  {Whitfield}}]{setia2020reducing}%
  \BibitemOpen
  \bibfield  {author} {\bibinfo {author} {\bibfnamefont {K.}~\bibnamefont
  {Setia}}, \bibinfo {author} {\bibfnamefont {R.}~\bibnamefont {Chen}},
  \bibinfo {author} {\bibfnamefont {J.~E.}\ \bibnamefont {Rice}}, \bibinfo
  {author} {\bibfnamefont {A.}~\bibnamefont {Mezzacapo}}, \bibinfo {author}
  {\bibfnamefont {M.}~\bibnamefont {Pistoia}},\ and\ \bibinfo {author}
  {\bibfnamefont {J.~D.}\ \bibnamefont {Whitfield}},\ }\bibfield  {title}
  {\bibinfo {title} {Reducing qubit requirements for quantum simulations using
  molecular point group symmetries},\ }\href@noop {} {\bibfield  {journal}
  {\bibinfo  {journal} {Journal of Chemical Theory and Computation}\ }\textbf
  {\bibinfo {volume} {16}},\ \bibinfo {pages} {6091} (\bibinfo {year}
  {2020})}\BibitemShut {NoStop}%
\bibitem [{\citenamefont {McClean}\ \emph {et~al.}(2020)\citenamefont
  {McClean}, \citenamefont {Rubin}, \citenamefont {Sung}, \citenamefont
  {Kivlichan}, \citenamefont {Bonet-Monroig}, \citenamefont {Cao},
  \citenamefont {Dai}, \citenamefont {Fried}, \citenamefont {Gidney},
  \citenamefont {Gimby} \emph {et~al.}}]{mcclean2020openfermion}%
  \BibitemOpen
  \bibfield  {author} {\bibinfo {author} {\bibfnamefont {J.~R.}\ \bibnamefont
  {McClean}}, \bibinfo {author} {\bibfnamefont {N.~C.}\ \bibnamefont {Rubin}},
  \bibinfo {author} {\bibfnamefont {K.~J.}\ \bibnamefont {Sung}}, \bibinfo
  {author} {\bibfnamefont {I.~D.}\ \bibnamefont {Kivlichan}}, \bibinfo {author}
  {\bibfnamefont {X.}~\bibnamefont {Bonet-Monroig}}, \bibinfo {author}
  {\bibfnamefont {Y.}~\bibnamefont {Cao}}, \bibinfo {author} {\bibfnamefont
  {C.}~\bibnamefont {Dai}}, \bibinfo {author} {\bibfnamefont {E.~S.}\
  \bibnamefont {Fried}}, \bibinfo {author} {\bibfnamefont {C.}~\bibnamefont
  {Gidney}}, \bibinfo {author} {\bibfnamefont {B.}~\bibnamefont {Gimby}}, \emph
  {et~al.},\ }\bibfield  {title} {\bibinfo {title} {Openfermion: the electronic
  structure package for quantum computers},\ }\href@noop {} {\bibfield
  {journal} {\bibinfo  {journal} {Quantum Science and Technology}\ }\textbf
  {\bibinfo {volume} {5}},\ \bibinfo {pages} {034014} (\bibinfo {year}
  {2020})}\BibitemShut {NoStop}%
\bibitem [{\citenamefont {Turney}\ \emph {et~al.}(2012)\citenamefont {Turney},
  \citenamefont {Simmonett}, \citenamefont {Parrish}, \citenamefont
  {Hohenstein}, \citenamefont {Evangelista}, \citenamefont {Fermann},
  \citenamefont {Mintz}, \citenamefont {Burns}, \citenamefont {Wilke},
  \citenamefont {Abrams} \emph {et~al.}}]{turney2012psi4}%
  \BibitemOpen
  \bibfield  {author} {\bibinfo {author} {\bibfnamefont {J.~M.}\ \bibnamefont
  {Turney}}, \bibinfo {author} {\bibfnamefont {A.~C.}\ \bibnamefont
  {Simmonett}}, \bibinfo {author} {\bibfnamefont {R.~M.}\ \bibnamefont
  {Parrish}}, \bibinfo {author} {\bibfnamefont {E.~G.}\ \bibnamefont
  {Hohenstein}}, \bibinfo {author} {\bibfnamefont {F.~A.}\ \bibnamefont
  {Evangelista}}, \bibinfo {author} {\bibfnamefont {J.~T.}\ \bibnamefont
  {Fermann}}, \bibinfo {author} {\bibfnamefont {B.~J.}\ \bibnamefont {Mintz}},
  \bibinfo {author} {\bibfnamefont {L.~A.}\ \bibnamefont {Burns}}, \bibinfo
  {author} {\bibfnamefont {J.~J.}\ \bibnamefont {Wilke}}, \bibinfo {author}
  {\bibfnamefont {M.~L.}\ \bibnamefont {Abrams}}, \emph {et~al.},\ }\bibfield
  {title} {\bibinfo {title} {Psi4: an open-source ab initio electronic
  structure program},\ }\href@noop {} {\bibfield  {journal} {\bibinfo
  {journal} {Wiley Interdisciplinary Reviews: Computational Molecular Science}\
  }\textbf {\bibinfo {volume} {2}},\ \bibinfo {pages} {556} (\bibinfo {year}
  {2012})}\BibitemShut {NoStop}%
\bibitem [{\citenamefont {Virtanen}\ \emph {et~al.}(2020)\citenamefont
  {Virtanen}, \citenamefont {Gommers}, \citenamefont {Oliphant}, \citenamefont
  {Haberland}, \citenamefont {Reddy}, \citenamefont {Cournapeau}, \citenamefont
  {Burovski}, \citenamefont {Peterson}, \citenamefont {Weckesser},
  \citenamefont {Bright}, \citenamefont {{van der Walt}}, \citenamefont
  {Brett}, \citenamefont {Wilson}, \citenamefont {Millman}, \citenamefont
  {Mayorov}, \citenamefont {Nelson}, \citenamefont {Jones}, \citenamefont
  {Kern}, \citenamefont {Larson}, \citenamefont {Carey}, \citenamefont {Polat},
  \citenamefont {Feng}, \citenamefont {Moore}, \citenamefont {{VanderPlas}},
  \citenamefont {Laxalde}, \citenamefont {Perktold}, \citenamefont {Cimrman},
  \citenamefont {Henriksen}, \citenamefont {Quintero}, \citenamefont {Harris},
  \citenamefont {Archibald}, \citenamefont {Ribeiro}, \citenamefont
  {Pedregosa}, \citenamefont {{van Mulbregt}},\ and\ \citenamefont {{SciPy 1.0
  Contributors}}}]{2020SciPy-NMeth}%
  \BibitemOpen
  \bibfield  {author} {\bibinfo {author} {\bibfnamefont {P.}~\bibnamefont
  {Virtanen}}, \bibinfo {author} {\bibfnamefont {R.}~\bibnamefont {Gommers}},
  \bibinfo {author} {\bibfnamefont {T.~E.}\ \bibnamefont {Oliphant}}, \bibinfo
  {author} {\bibfnamefont {M.}~\bibnamefont {Haberland}}, \bibinfo {author}
  {\bibfnamefont {T.}~\bibnamefont {Reddy}}, \bibinfo {author} {\bibfnamefont
  {D.}~\bibnamefont {Cournapeau}}, \bibinfo {author} {\bibfnamefont
  {E.}~\bibnamefont {Burovski}}, \bibinfo {author} {\bibfnamefont
  {P.}~\bibnamefont {Peterson}}, \bibinfo {author} {\bibfnamefont
  {W.}~\bibnamefont {Weckesser}}, \bibinfo {author} {\bibfnamefont
  {J.}~\bibnamefont {Bright}}, \bibinfo {author} {\bibfnamefont {S.~J.}\
  \bibnamefont {{van der Walt}}}, \bibinfo {author} {\bibfnamefont
  {M.}~\bibnamefont {Brett}}, \bibinfo {author} {\bibfnamefont
  {J.}~\bibnamefont {Wilson}}, \bibinfo {author} {\bibfnamefont {K.~J.}\
  \bibnamefont {Millman}}, \bibinfo {author} {\bibfnamefont {N.}~\bibnamefont
  {Mayorov}}, \bibinfo {author} {\bibfnamefont {A.~R.~J.}\ \bibnamefont
  {Nelson}}, \bibinfo {author} {\bibfnamefont {E.}~\bibnamefont {Jones}},
  \bibinfo {author} {\bibfnamefont {R.}~\bibnamefont {Kern}}, \bibinfo {author}
  {\bibfnamefont {E.}~\bibnamefont {Larson}}, \bibinfo {author} {\bibfnamefont
  {C.~J.}\ \bibnamefont {Carey}}, \bibinfo {author} {\bibfnamefont
  {{\.I}.}~\bibnamefont {Polat}}, \bibinfo {author} {\bibfnamefont
  {Y.}~\bibnamefont {Feng}}, \bibinfo {author} {\bibfnamefont {E.~W.}\
  \bibnamefont {Moore}}, \bibinfo {author} {\bibfnamefont {J.}~\bibnamefont
  {{VanderPlas}}}, \bibinfo {author} {\bibfnamefont {D.}~\bibnamefont
  {Laxalde}}, \bibinfo {author} {\bibfnamefont {J.}~\bibnamefont {Perktold}},
  \bibinfo {author} {\bibfnamefont {R.}~\bibnamefont {Cimrman}}, \bibinfo
  {author} {\bibfnamefont {I.}~\bibnamefont {Henriksen}}, \bibinfo {author}
  {\bibfnamefont {E.~A.}\ \bibnamefont {Quintero}}, \bibinfo {author}
  {\bibfnamefont {C.~R.}\ \bibnamefont {Harris}}, \bibinfo {author}
  {\bibfnamefont {A.~M.}\ \bibnamefont {Archibald}}, \bibinfo {author}
  {\bibfnamefont {A.~H.}\ \bibnamefont {Ribeiro}}, \bibinfo {author}
  {\bibfnamefont {F.}~\bibnamefont {Pedregosa}}, \bibinfo {author}
  {\bibfnamefont {P.}~\bibnamefont {{van Mulbregt}}},\ and\ \bibinfo {author}
  {\bibnamefont {{SciPy 1.0 Contributors}}},\ }\bibfield  {title} {\bibinfo
  {title} {{{SciPy} 1.0: Fundamental Algorithms for Scientific Computing in
  Python}},\ }\href {https://doi.org/10.1038/s41592-019-0686-2} {\bibfield
  {journal} {\bibinfo  {journal} {Nature Methods}\ }\textbf {\bibinfo {volume}
  {17}},\ \bibinfo {pages} {261} (\bibinfo {year} {2020})}\BibitemShut
  {NoStop}%
\bibitem [{Note1()}]{Note1}%
  \BibitemOpen
  \bibinfo {note} {\protect \url
  {https://cccbdb.nist.gov/energy3x.asp?method=15\&basis=6\&charge=0}}\BibitemShut
  {NoStop}%
\bibitem [{\citenamefont {Hong}\ \emph {et~al.}(2022)\citenamefont {Hong},
  \citenamefont {Tsai}, \citenamefont {Chou}, \citenamefont {Chen},
  \citenamefont {Tsai}, \citenamefont {Chen}, \citenamefont {Kuo},
  \citenamefont {Srolovitz}, \citenamefont {Hu}, \citenamefont {Cheng} \emph
  {et~al.}}]{hong2022accurate}%
  \BibitemOpen
  \bibfield  {author} {\bibinfo {author} {\bibfnamefont {C.-L.}\ \bibnamefont
  {Hong}}, \bibinfo {author} {\bibfnamefont {T.}~\bibnamefont {Tsai}}, \bibinfo
  {author} {\bibfnamefont {J.-P.}\ \bibnamefont {Chou}}, \bibinfo {author}
  {\bibfnamefont {P.-J.}\ \bibnamefont {Chen}}, \bibinfo {author}
  {\bibfnamefont {P.-K.}\ \bibnamefont {Tsai}}, \bibinfo {author}
  {\bibfnamefont {Y.-C.}\ \bibnamefont {Chen}}, \bibinfo {author}
  {\bibfnamefont {E.-J.}\ \bibnamefont {Kuo}}, \bibinfo {author} {\bibfnamefont
  {D.}~\bibnamefont {Srolovitz}}, \bibinfo {author} {\bibfnamefont
  {A.}~\bibnamefont {Hu}}, \bibinfo {author} {\bibfnamefont {Y.-C.}\
  \bibnamefont {Cheng}}, \emph {et~al.},\ }\bibfield  {title} {\bibinfo {title}
  {Accurate and efficient quantum computations of molecular properties using
  daubechies wavelet molecular orbitals: A benchmark study against experimental
  data},\ }\href@noop {} {\bibfield  {journal} {\bibinfo  {journal} {arXiv
  preprint arXiv:2205.14476}\ } (\bibinfo {year} {2022})}\BibitemShut {NoStop}%
\bibitem [{\citenamefont {Colless}\ \emph {et~al.}(2018)\citenamefont
  {Colless}, \citenamefont {Ramasesh}, \citenamefont {Dahlen}, \citenamefont
  {Blok}, \citenamefont {Kimchi-Schwartz}, \citenamefont {McClean},
  \citenamefont {Carter}, \citenamefont {de~Jong},\ and\ \citenamefont
  {Siddiqi}}]{colless2018computation}%
  \BibitemOpen
  \bibfield  {author} {\bibinfo {author} {\bibfnamefont {J.~I.}\ \bibnamefont
  {Colless}}, \bibinfo {author} {\bibfnamefont {V.~V.}\ \bibnamefont
  {Ramasesh}}, \bibinfo {author} {\bibfnamefont {D.}~\bibnamefont {Dahlen}},
  \bibinfo {author} {\bibfnamefont {M.~S.}\ \bibnamefont {Blok}}, \bibinfo
  {author} {\bibfnamefont {M.~E.}\ \bibnamefont {Kimchi-Schwartz}}, \bibinfo
  {author} {\bibfnamefont {J.~R.}\ \bibnamefont {McClean}}, \bibinfo {author}
  {\bibfnamefont {J.}~\bibnamefont {Carter}}, \bibinfo {author} {\bibfnamefont
  {W.~A.}\ \bibnamefont {de~Jong}},\ and\ \bibinfo {author} {\bibfnamefont
  {I.}~\bibnamefont {Siddiqi}},\ }\bibfield  {title} {\bibinfo {title}
  {Computation of molecular spectra on a quantum processor with an
  error-resilient algorithm},\ }\href@noop {} {\bibfield  {journal} {\bibinfo
  {journal} {Physical Review X}\ }\textbf {\bibinfo {volume} {8}},\ \bibinfo
  {pages} {011021} (\bibinfo {year} {2018})}\BibitemShut {NoStop}%
\bibitem [{\citenamefont {Higgott}\ \emph {et~al.}(2019)\citenamefont
  {Higgott}, \citenamefont {Wang},\ and\ \citenamefont
  {Brierley}}]{higgott2019variational}%
  \BibitemOpen
  \bibfield  {author} {\bibinfo {author} {\bibfnamefont {O.}~\bibnamefont
  {Higgott}}, \bibinfo {author} {\bibfnamefont {D.}~\bibnamefont {Wang}},\ and\
  \bibinfo {author} {\bibfnamefont {S.}~\bibnamefont {Brierley}},\ }\bibfield
  {title} {\bibinfo {title} {Variational quantum computation of excited
  states},\ }\href@noop {} {\bibfield  {journal} {\bibinfo  {journal}
  {Quantum}\ }\textbf {\bibinfo {volume} {3}},\ \bibinfo {pages} {156}
  (\bibinfo {year} {2019})}\BibitemShut {NoStop}%
\bibitem [{\citenamefont {O’Malley}\ \emph {et~al.}(2016)\citenamefont
  {O’Malley}, \citenamefont {Babbush}, \citenamefont {Kivlichan},
  \citenamefont {Romero}, \citenamefont {McClean}, \citenamefont {Barends},
  \citenamefont {Kelly}, \citenamefont {Roushan}, \citenamefont {Tranter},
  \citenamefont {Ding} \emph {et~al.}}]{o2016scalable}%
  \BibitemOpen
  \bibfield  {author} {\bibinfo {author} {\bibfnamefont {P.~J.}\ \bibnamefont
  {O’Malley}}, \bibinfo {author} {\bibfnamefont {R.}~\bibnamefont {Babbush}},
  \bibinfo {author} {\bibfnamefont {I.~D.}\ \bibnamefont {Kivlichan}}, \bibinfo
  {author} {\bibfnamefont {J.}~\bibnamefont {Romero}}, \bibinfo {author}
  {\bibfnamefont {J.~R.}\ \bibnamefont {McClean}}, \bibinfo {author}
  {\bibfnamefont {R.}~\bibnamefont {Barends}}, \bibinfo {author} {\bibfnamefont
  {J.}~\bibnamefont {Kelly}}, \bibinfo {author} {\bibfnamefont
  {P.}~\bibnamefont {Roushan}}, \bibinfo {author} {\bibfnamefont
  {A.}~\bibnamefont {Tranter}}, \bibinfo {author} {\bibfnamefont
  {N.}~\bibnamefont {Ding}}, \emph {et~al.},\ }\bibfield  {title} {\bibinfo
  {title} {Scalable quantum simulation of molecular energies},\ }\href@noop {}
  {\bibfield  {journal} {\bibinfo  {journal} {Physical Review X}\ }\textbf
  {\bibinfo {volume} {6}},\ \bibinfo {pages} {031007} (\bibinfo {year}
  {2016})}\BibitemShut {NoStop}%
\bibitem [{\citenamefont {Bergholm}\ \emph {et~al.}(2018)\citenamefont
  {Bergholm}, \citenamefont {Izaac}, \citenamefont {Schuld}, \citenamefont
  {Gogolin}, \citenamefont {Alam}, \citenamefont {Ahmed}, \citenamefont
  {Arrazola}, \citenamefont {Blank}, \citenamefont {Delgado}, \citenamefont
  {Jahangiri} \emph {et~al.}}]{bergholm2018pennylane}%
  \BibitemOpen
  \bibfield  {author} {\bibinfo {author} {\bibfnamefont {V.}~\bibnamefont
  {Bergholm}}, \bibinfo {author} {\bibfnamefont {J.}~\bibnamefont {Izaac}},
  \bibinfo {author} {\bibfnamefont {M.}~\bibnamefont {Schuld}}, \bibinfo
  {author} {\bibfnamefont {C.}~\bibnamefont {Gogolin}}, \bibinfo {author}
  {\bibfnamefont {M.~S.}\ \bibnamefont {Alam}}, \bibinfo {author}
  {\bibfnamefont {S.}~\bibnamefont {Ahmed}}, \bibinfo {author} {\bibfnamefont
  {J.~M.}\ \bibnamefont {Arrazola}}, \bibinfo {author} {\bibfnamefont
  {C.}~\bibnamefont {Blank}}, \bibinfo {author} {\bibfnamefont
  {A.}~\bibnamefont {Delgado}}, \bibinfo {author} {\bibfnamefont
  {S.}~\bibnamefont {Jahangiri}}, \emph {et~al.},\ }\bibfield  {title}
  {\bibinfo {title} {Pennylane: Automatic differentiation of hybrid
  quantum-classical computations},\ }\href@noop {} {\bibfield  {journal}
  {\bibinfo  {journal} {arXiv preprint arXiv:1811.04968}\ } (\bibinfo {year}
  {2018})}\BibitemShut {NoStop}%
\bibitem [{\citenamefont {van~de Wetering}(2020)}]{van2020zx}%
  \BibitemOpen
  \bibfield  {author} {\bibinfo {author} {\bibfnamefont {J.}~\bibnamefont
  {van~de Wetering}},\ }\bibfield  {title} {\bibinfo {title} {Zx-calculus for
  the working quantum computer scientist},\ }\href@noop {} {\bibfield
  {journal} {\bibinfo  {journal} {arXiv preprint arXiv:2012.13966}\ } (\bibinfo
  {year} {2020})}\BibitemShut {NoStop}%
\bibitem [{\citenamefont {Coecke}\ \emph {et~al.}(2021)\citenamefont {Coecke},
  \citenamefont {Horsman}, \citenamefont {Kissinger},\ and\ \citenamefont
  {Wang}}]{coecke2021kindergarden}%
  \BibitemOpen
  \bibfield  {author} {\bibinfo {author} {\bibfnamefont {B.}~\bibnamefont
  {Coecke}}, \bibinfo {author} {\bibfnamefont {D.}~\bibnamefont {Horsman}},
  \bibinfo {author} {\bibfnamefont {A.}~\bibnamefont {Kissinger}},\ and\
  \bibinfo {author} {\bibfnamefont {Q.}~\bibnamefont {Wang}},\ }\bibfield
  {title} {\bibinfo {title} {Kindergarden quantum mechanics graduates (... or
  how i learned to stop gluing lego together and love the zx-calculus)},\
  }\href@noop {} {\bibfield  {journal} {\bibinfo  {journal} {arXiv preprint
  arXiv:2102.10984}\ } (\bibinfo {year} {2021})}\BibitemShut {NoStop}%
\end{thebibliography}%

\end{document}